\documentclass[11pt,letterpaper]{amsart}

\usepackage{hyperref}
\usepackage{amssymb}
\usepackage{mathtools}
\usepackage{todonotes}
\usepackage{mathrsfs}

\newtheorem{Thm}{Theorem}

\newenvironment{Proof}{\begin{proof}}{\end{proof}}

\newcommand{\per}{\ensuremath{\text{per}\,}}
\newcommand{\R}{\ensuremath{\mathscr{R}}}
\newcommand{\Z}{\ensuremath{\mathbb{Z}}}
\newcommand{\F}{\ensuremath{\mathbb{F}}}
\newcommand{\bra}{\ensuremath{\langle}}
\newcommand{\ket}{\ensuremath{\rangle}}

\newcommand\multiset[2]%
{\mathchoice{\left(\kern-0.5em{\binom{#1}{#2}}\kern-0.5em\right)}
            {\bigl(\kern-0.3em{\binom{#1}{#2}}\kern-0.3em\bigr)}
            {\bigl(\kern-0.3em{\binom{#1}{#2}}\kern-0.3em\bigr)}
            {\bigl(\kern-0.3em{\binom{#1}{#2}}\kern-0.3em\bigr)}}

\begin{document}

\title[]{How Proofs are Prepared at Camelot*}

\thanks{*The research leading to these results has received funding from the Swedish Research Council grant VR 2012-4730 "Exact 
Exponential-Time Algorithms" (A.B.) and the 
European Research Council under the European Union's Seventh Framework 
Programme (FP/2007-2013) / ERC Grant Agreement 338077 ``Theory and 
Practice of Advanced Search and Enumeration'' (P.K.).
Work done in part while the authors were 
visiting the Simons Institute for the Theory of Computing.}

\author[]{Andreas Bj\"orklund \and Petteri Kaski}

\begin{abstract}
We study a design framework for robust, independently verifiable, 
and workload-balanced distributed algorithms working on a common 
input. An algorithm based on the framework is essentially 
a {\em distributed encoding procedure} for a Reed--Solomon code 
(each compute node is tasked to produce one or more symbols of 
the codeword), which enables (a) robustness against byzantine 
adversarial failures with intrinsic error-correction and 
identification of failed nodes, and (b) independent randomized 
verification to check the entire computation for correctness, 
which takes essentially no more resources than each node individually 
contributes to execute the computation. The framework also enables 
smooth tradeoffs between per-node compute time and the number of 
nodes used for computation.

The framework builds on recent {\em Merlin--Arthur}
proofs of batch evaluation of 
Williams~[{\em Electron.\ Colloq.\ Comput.\ Complexity}, 
Report TR16-002, January 2016] 
with the basic observation that 
{\em Merlin's magic is not needed} for batch evaluation---mere 
{\em Knights} can prepare the independently verifiable {\em proof}, 
in parallel, and with intrinsic error-correction.
Dually, {\em should Merlin materialize},
he can relieve the Knights and instantaneously supply the proof, 
in which case these algorithms are, {\em as is}, Merlin--Arthur 
protocols. 

The contribution of this paper is to show that in many cases 
the verifiable batch evaluation framework admits 
algorithms that match in total resource consumption the best 
known sequential algorithm for {\em solving} the problem. 
As our main technical result, we show that the $k$-cliques 
in an $n$-vertex graph can be counted {\em and} verified in per-node 
$O(n^{(\omega+\epsilon)k/6})$ time and space on 
$O(n^{(\omega+\epsilon)k/6})$ compute nodes, for any constant $\epsilon>0$
and positive integer $k$ divisible by $6$, 
where $2\leq\omega<2.3728639$ is the exponent 
of square matrix multiplication over the integers. This matches in 
total running time the best known sequential algorithm, due to 
Ne{\v{s}}et{\v{r}}il and Poljak 
[{\em Comment.~Math.~Univ.~Carolin.}~26 (1985) 415--419], and 
considerably improves its space usage and parallelizability.
Further results include novel algorithms for counting triangles 
in sparse graphs, computing the chromatic polynomial of a graph, 
and computing the Tutte polynomial of a graph.

\end{abstract}

\maketitle

\thispagestyle{empty}

\setcounter{page}{0}

\clearpage

\section{Introduction}

\subsection{A scene of distress and relief at Camelot}

Picture $K$ equally capable Knights seated around the Round Table,
distressed. At the center of the table stands the Input. 
The Knights have been tasked to prepare a Proof about the virtues of 
the input, and to make {\em extreme} haste: 
a proof must be prepared in time
$T/K$, where $T$ is the {\em fastest time known in all of Britannia} 
for any single soul to {\em reveal} such subtle virtues, 
let alone give {\em uncontestable proof} thereof. 

Nigh impossible is the task in fact, for the Lady Morgana has 
enchanted many a poor Knight with her cunning dark magic, yet virtues 
must be revealed and proof thereof prepared. And not just any proof, 
but one that any lone soul can check, in time $T/K$ and with a 
few tosses of a fair coin, and next to never be convinced if foul 
virtues it claims. 

However, not all is lost, for the Knights recall the teachings of
the wizard Merlin%
\footnote{Alas, the good wizard is taking a vacation, hence the distress
at Camelot.}
and his powerful proofs that King Arthur so treasures.
So the Knights agree to together evaluate in batch, 
and to individually decode-and-check for signs of the darkest 
of magic...

\subsection{A template for community computation over common input}

Building on Merlin--Arthur proofs~\cite{Babai1988} of 
{\em batch evaluation} of Williams~\cite{Williams2016}, 
this paper documents a ``Camelot template'' 
for verifiable distributed computation that is robust against 
{\em adversarial byzantine failures} at the nodes and produces 
a static, independently verifiable {\em proof} that the computation 
succeeded. In essence, we observe that the act of {\em preparing}
a proof of batch evaluation in the Williams framework is 
{\em magicless}%
\footnote{In Arthurian parlance, 
{\em Merlin's magic is not needed}---{\em mere Knights}
prepare the {\em proof}, in parallel, and with intrinsic 
error-correction.
Dually, {\em should Merlin materialize},
he can relieve the Knights and instantaneously supply the proof, 
in which case these algorithms are, {\em as is}, Merlin--Arthur 
protocols.} and {\em robust} against errors. Our contribution
is in particular to show that for a number of problems 
such robust and verifiable algorithms essentially match in 
total resource consumption the best known sequential 
algorithm for {\em solving} the problem.

By robustness we mean here that during proof preparation 
a substantial number%
\footnote{In precise terms, up to the decoding limit of the
decoder used for the Reed--Solomon code, such as 
at most $(e-d-1)/2$ nodes if we use, e.g.,~the Gao~\cite{Gao2003} 
decoder and $e$ evaluation points for a proof polynomial of degree 
$d$; see~\S\ref{sect:encode-decode}.}{}
of nodes may undergo byzantine failure and a correct proof will 
still be produced at each correctly functioning node.
(We refer to e.g.~Tiwari {\em et al.}~\cite{Tiwari2015} for 
recent empirical motivation why robustness is an increasingly 
desirable goal in algorithm design.)

By verifiability we mean here that 
if the computation {\em did} fail, each node detects this individually
{\em regardless of how many nodes experienced byzantine failure}
during computation. In more precise terms, each node (or other 
entity) can verify the static proof against the input so that the 
probability of accepting an incorrect proof is negligible {\em and} 
controllable by the verifier {\em after} the proof is presented
for verification. Furthermore, the resources needed 
for verification are essentially identical to the resources 
that a single node contributes to the community.

\subsection{How Camelot prepares, corrects, and checks a proof}

Let us interpret the Williams batch evaluation 
framework~\cite{Williams2016} as a template for distributed 
computing with $K$ compute nodes. 
The key idea is to task the $K$ nodes to 
make {\em evaluations} of a univariate polynomial 
\[
P(x)=p_0+p_1x+p_2x^2+\ldots+p_dx^d\pmod q\,,
\]
where $q$ is a prime that we assume each node can easily compute 
by looking at the common input.%
\footnote{For ease of exposition, we work over fields of prime order 
though generalizations to field extensions are possible, e.g.,
to obtain better fault tolerance. Further dimensions for 
generalizations include, e.g., the use of multivariate 
(Reed--Muller~\cite{Muller1954,Reed1954}) polynomial codes.}{}
(We also assume that each node can easily compute an upper bound 
for $d$ from the common input.)

The polynomial $P(x)$ has been carefully constructed so that  
(a) it reveals%
\footnote{Possibly requiring iteration and postprocessing, e.g.,
the use of multiple distinct primes $q$ and the Chinese Remainder 
Theorem to reconstruct large integers.}{}
the desired properties of the input
{\em and} 
(b) the {\em coefficients} $p_0,p_1,\ldots,p_d\in\Z_q$ 
constitute a probabilistically verifiable {\em proof} that these 
properties have been correctly computed. 
In fact, the latter property is {\em intrinsic} to the template.

\medskip
\noindent
{\em 1. Proof preparation, in distributed encoded form.} 
We assume that the input (the problem instance) is available at
each of the $K$ nodes; for example, the input may be broadcast
over a network to all the nodes.
For a prime $q$ and an integer $e$ with $d+1\leq e\leq q$, 
let us collectively task the $K$ nodes to compute the sequence 
of $e$ evaluations 
\begin{equation}
\label{eq:poly-eval}
P(0),P(1),\ldots,P(e-1) \pmod q
\end{equation}
so that each node is responsible for about $e/K$ evaluations,
and broadcasts its evaluations to the other nodes once they are 
ready.%
\footnote{In particular, in total $e$ integers modulo $q$ are
broadcast over the network, with the objective that all nodes
receive the $e$ evaluations. (If only a single node is to 
error-correct and verify the proof, then no broadcasting is
necessary in this step -- each node forwards its evaluations only 
to the verifier node.)}{}
This sequence of evaluations is, {\em by definition}, an
encoding of the proof in the classical nonsystematic
{\em Reed--Solomon code}~\cite{Reed1960} 
(see \S\ref{sect:preliminaries}).
From the perspective of {\em community computation}, this is a 
serendipitous property:

\medskip
\noindent
{\em 2. Error-correction during preparation of the proof.} 
It is possible to error-correct a proof {\em in preparation} to cope 
with failing nodes in the community of nodes, up to the 
error-correction capability of the code and the decoding algorithm. 
Indeed, since the nodes are tasked to produce entries in 
a Reed--Solomon codeword, each node can {\em individually}
run a {\em fast} Reed--Solomon decoder~\cite{Gao2003} 
(see \S\ref{sect:preliminaries})
on the entries it has received from the community to recover
{\em both} the actual proof $p_0,p_1,\ldots,p_d\in\Z_q$ {\em and} 
the evaluations $P(0),P(1),\ldots,P(e-1)\in\Z_q$ {\em that failed}, 
that is, the {\em error locations} identified by the decoder.
Thus, if a node so desires, it can, on its own, identify the 
nodes that did not properly participate in the community effort
by looking at the error locations.
Furthermore,%
\footnote{In precise terms, a collective conclusion can be reached
{\em assuming the nodes agree on the evaluations} 
$P(0),P(1),\ldots,P(e-1)\in\Z_q$ that were produced by the community.
It is perhaps a realistic practical assumption that such an 
agreement can be reached by an appropriate broadcast mechanism for
informing all the nodes about a node's contributions. We observe that
such collective agreement {\em need, however, not be available for purposes 
of decoding the proof}---the decoder works as long as it receives
a sufficient number of correct entries of the codeword, and these
entries may be different for different nodes.}
all nodes can reach the same {\em collective} conclusion about 
the community contribution on their own.

\medskip
\noindent
{\em 3. Checking a putative proof for correctness.} 
Each node (or any other entity) with access to the common input
and to a sequence $\tilde p_0,\tilde p_1,\ldots,\tilde p_d\in\Z_q$ 
can, probabilistically, check whether the sequence is in fact the 
correct proof $p_0,p_1,\ldots,p_d\in\Z_q$. 
All a node has to do is to select a uniform random integer 
$x_0\in\Z_q$ and accept the proof if and only if
\begin{equation}
\label{eq:verify}
P(x_0)=\sum_{j=0}^d \tilde p_j x_0^j\pmod q 
\end{equation}
holds, at the cost of one evaluation of $P(x)$. (That is, the verifier
executes the same algorithm that the nodes use for preparing the proof.%
\footnote{In precise terms, this is to obtain the left-hand side 
of \eqref{eq:verify}. The right-hand side can be computed from
$\tilde p_0,\tilde p_1,\ldots,\tilde p_d\in\Z_q$ using Horner's rule.}{})
If the sequence $\tilde p_0,\tilde p_1,\ldots,\tilde p_d$ {\em is} the 
correct proof, the verifier always accepts. If the sequence is incorrect, 
it follows immediately from the fundamental theorem of algebra
that $P(x_0)\neq\sum_{j=0}^d \tilde p_j x_0^j\pmod q$ holds with
probability at least $1-d/q$. Furthermore, since the verifier knows
$q$ and an upper bound for $d$, the verifier can control the 
probability of accepting an incorrect proof by independent 
repetitions of the test.

\subsection{Optimality relative to best known sequential algorithms}

Ideally, a Camelot algorithm should match in {\em total}%
\footnote{The sum of running times of all the nodes.}{}
running time the worst-case running time of the best known 
{\em sequential} algorithm for {\em solving} the problem. 
By design, such a Camelot algorithm not only solves the problem, 
but does so in {\em parallel}, 
{\em robustly}, and gives {\em an independently verifiable proof} 
that it succeeded. What is more, the parallel part of the computation 
\eqref{eq:poly-eval} 
consists of evaluating {\em the same} polynomial $P$ at distinct 
points, making the parallel computations intrinsically 
{\em workload-balanced} and ideal for {\em vector-parallelization} 
({\em single-instruction multiple-data (SIMD) parallelization})
since each evaluation follows the same instruction stream.

In essence, we seek the following {\em optimal speedup tradeoff} 
for a Camelot design.
Assume that the best known sequential algorithm solves the problem 
essentially
in time $T$. Then, an optimal Camelot design 
for $K$ nodes should run in $E=T/K$ proof construction time and space 
for each node, and hence $E$ evaluation (proof verification) time 
for a proof of size $K$, for any $K\leq T^{1/2}$.
Here the upper bound $T^{1/2}$ is intrinsic to the framework 
since each node receives the whole proof for individual decoding 
and hence $E\geq K$.% 
\footnote{This of course does not rule out other frameworks
where a single node does not work with the entire proof.}{}

In practice, our subsequent algorithm designs will not attain the 
optimal tradeoff for all problems considered. 
{\em However, for a number of canonical problems we are able 
to attain the optimal tradeoff.}

\subsection{Our results---four highlights}

This section summarizes the four main technical results of this paper
that highlight the technical versatility of the Camelot framework.
For ease of exposition we present our results as non-tradeoff versions 
with the maximum value of $K$ (and hence the fastest wall-clock runtime $E$ 
to prepare and check the proof) that we can attain, with the understanding 
that a smooth tradeoff up to the maximum $K$ is possible and can be read off
the proofs of our results. 

When stating our results, we refer to $K$ as the {\em proof size}
and $E$ as the {\em (running) time} of a Camelot algorithm.
Put otherwise, the time $E$ of a Camelot algorithm is the {\em wall-clock} 
time to prepare and verify the proof using $K$ compute nodes working in 
parallel on the common input. The {\em total time} used by all the nodes 
is $EK$. The underlying precise degree $d$ of the proof polynomial $P(x)$ 
is, up to factors suppressed by asymptotic notation, identically bounded 
as the stated proof size.%
\footnote{%
The parameters $e$ and $q$ can be chosen 
with similar asymptotic bounds as $d$. We choose to omit the details of
these parameters to keep the statement of our results concise, please 
consult the proofs for precise bounds on $d$ to enable fine-tuning of 
selection of $d+1\leq e\leq q$ as appropriate for desired error-tolerance
(during proof preparation) and soundness guarantee (for proof verification).}{}
With asymptotic notation we follow the convention 
that $\tilde O(\cdot)$ suppresses a factor polylogarithmic 
in the input size and $O^*(\cdot)$ suppresses a factor 
polynomial in the input size.

Each Camelot algorithm defines, {\em as is}, a {\em Merlin--Arthur} 
protocol \cite{Babai1988}, where Merlin's proof has the stated size,
and Arthur's verification time is the stated time for the Camelot
algorithm. 

\medskip
\noindent
{\em 1. Optimal Camelot algorithms in polynomial time: counting small cliques.}
Our main result is that the problem of counting all the $k$-cliques in 
a graph admits a Camelot algorithm {\em that attains the optimal tradeoff
relative to the best known sequential algorithm}.
Let us write $2\leq\omega<2.3728639$ for the exponent of square matrix
multiplication over the integers~\cite{LeGall2014}.

\begin{Thm}
\label{thm:k-clique-camelot}
For any constant $\epsilon>0$ and any positive integer $k$
divisible by $6$, the number of\/ $k$-cliques in an $n$-vertex graph
can be computed with a Camelot algorithm that constructs 
a proof of size $O\bigl(n^{(\omega+\epsilon)k/6}\bigr)$ 
in time $O\bigl(n^{(\omega+\epsilon)k/6}\bigr)$.
\end{Thm}

The best known sequential algorithm, due to 
Ne{\v{s}}et{\v{r}}il and Poljak \cite{Nesetril1985}
(see also Eisenbrand and Grandoni~\cite{Eisenbrand2004}),
runs in time $O(n^{(\omega+\epsilon)k/3})$ and space $O(n^{2k/3})$.
In particular, Theorem~\ref{thm:k-clique-camelot} gives an optimal tradeoff
and Theorem~\ref{thm:k-clique} improves the space usage and parallelizability
of the best known sequential algorithm.

The algorithm in Theorem~\ref{thm:k-clique-camelot} is based 
on a novel, more space-efficient family of arithmetic circuits for 
evaluating a particular $\binom{6}{2}$-linear form that enables a 
very space-efficient design for counting 6-cliques in a graph. 
We state the latter result independently of the Camelot framework:

\begin{Thm}
\label{thm:k-clique}
For any constant $\epsilon>0$ and any positive integer $k$
divisible by $6$, there exists an algorithm that counts 
the $k$-cliques in a given
$n$-vertex graph in time $O\bigl(n^{(\omega+\epsilon)k/6}\bigr)$ 
and space $O\bigl(n^{k/3}\bigr)$.
Furthermore, the algorithm can be executed in per-node time
$O\bigl(n^{(\omega+\epsilon)k/6}\bigr)$ and space 
$O\bigl(n^{k/3}\bigr)$ on $O\bigl(n^{(\omega+\epsilon)k/6}\bigr)$ 
parallel compute nodes.
\end{Thm}

Observe in particular that for $k=6$ the space usage
in Theorem~\ref{thm:k-clique} is $O(n^2)$, that is, the 
algorithm uses asymptotically no more space than the
input graph.

\medskip
\noindent
{\em 2. Sparsity-aware Camelot algorithms: counting triangles.}
As our second highlight we show that Camelot algorithms can be 
made aware of {\em sparsity} in the given input. 
In particular, we study the task of counting triangles in 
a given $n$-vertex graph. Taking the 
$O(n^{\omega+\epsilon})$-time sequential algorithm of 
Itai and Rodeh~\cite{Itai1978} as our baseline for parallelization,
we present a Camelot algorithm whose running time is essentially 
linear in the size of the input.

\begin{Thm}
\label{thm:triangle-camelot}
For any constant $\epsilon>0$, the number of\/ triangles 
in an $n$-vertex graph with $m\geq n^{(\omega+\epsilon)/2}$ edges 
can be computed with a Camelot algorithm that constructs 
a proof of size $O\bigl(n^{\omega+\epsilon}/m\bigr)$ 
in time $\tilde O\bigl(m\bigr)$.
\end{Thm}

The algorithm in Theorem~\ref{thm:triangle-camelot} is based 
on a novel, more space-efficient split/sparse polynomial 
extension of Yates's algorithm~\cite{Yates1937} for 
computing matrix--vector products when the matrix has 
iterated Kronecker product structure.
We state a space-efficient and parallel version of the
result independently of the Camelot framework:

\begin{Thm}
\label{thm:triangle}
For any constant $\epsilon>0$, there exists an algorithm 
that counts the triangles in a given 
connected $n$-vertex graph with $m$ edges 
in time $O\bigl(n^{\omega+\epsilon})$ and space 
$\tilde O\bigl(m\bigr)$.
Furthermore, the algorithm can be executed in per-node time
and space $\tilde O\bigl(m\bigr)$ on 
$O\bigl(n^{\omega+\epsilon}/m\bigr)$ parallel compute nodes.
\end{Thm}

\noindent
{\em Remark.}
For a {\em sparse} input consisting of $m$ edges, the fastest 
known sequential algorithm for triangle counting
due to Alon, Yuster, and Zwick~\cite{Alon1997} runs in time 
$O\bigl(m^{2(\omega+\epsilon)/(\omega+1)}\bigr)$. Our designs in 
Theorem~\ref{thm:triangle-camelot} and Theorem~\ref{thm:triangle}
only match the weaker Itai--Rodeh bound $O(n^{\omega+\epsilon})$;
matching the Alon--Yuster--Zwick bound with an optimal 
Camelot algorithm remains an open problem.

\medskip
If we seek only a parallel algorithm, the split/sparse variant
of Yates's algorithm enables us to match the Alon--Yuster--Zwick
bound:

\begin{Thm}
\label{thm:triangle-sparse}
For any constant $\epsilon>0$, there exists an algorithm 
that counts the triangles in a given graph with $m$ edges 
in time $O\bigl(m^{2(\omega+\epsilon)/(\omega+1)})$ and space 
$\tilde O\bigl(m\bigr)$.
Furthermore, the algorithm can be executed in per-node time
and space $\tilde O\bigl(m\bigr)$ on 
$O\bigl(m^{(\omega-1+\epsilon)/(\omega+1)}\bigr)$ 
parallel compute nodes.
\end{Thm}

\medskip
\noindent
{\em 3. Optimal Camelot algorithms in exponential time: 
the chromatic polynomial.}
As our third highlight result, we show that the problem of computing
the chromatic polynomial of a given graph admits an optimal 
Camelot algorithm. 

\begin{Thm}
\label{thm:chromatic-polynomial}
The chromatic polynomial of an $n$-vertex graph can be computed
with a Camelot algorithm that constructs a proof of
size $O^*(2^{n/2})$ in time $O^*(2^{n/2})$.
\end{Thm}

Here the best known sequential algorithm runs in $O^*(2^n)$ time and
$O^*(1.292^n)$ space \cite{BHK2009,BHKK2011}. In particular, 
Theorem~\ref{thm:chromatic-polynomial} gives an optimal tradeoff.

\medskip
\noindent
{\em 4. Near-optimal Camelot algorithms in exponential time: 
the Tutte polynomial.} 
The {\em Tutte polynomial} or the {\em bichromatic polynomial}
of a graph is a universal invariant for all
constant-coefficient deletion--contraction recurrences. 
The Tutte polynomial subsumes a large number of \#P-hard 
counting problems, such as the chromatic polynomial, the flow polynomial, 
the all-terminal reliability polynomial, the partition functions of the 
Ising and Potts models in statistical physics, the Jones polynomial 
of a link in knot theory, and more. 
(We refer to Welsh~\cite{Welsh1993} and Godsil and Royle~%
\cite{Godsil2001} for a detailed discussion on polynomial
graph invariants and universality.)

For the Tutte polynomial, we can almost attain the 
optimal tradeoff:%

\begin{Thm}
\label{thm:tutte-polynomial}
For any constant $\epsilon>0$, 
the Tutte polynomial of an $n$-vertex graph can be computed
with a Camelot algorithm that constructs a proof of
size $O^*(2^{n/3})$ in time $O^*(2^{(\omega+\epsilon)n/3})$ 
and per-node space $O^*(2^{2n/3})$.
\end{Thm}

Here the main bottleneck is the {\em space usage} of the best known 
sequential algorithm~\cite{BHKK2008}, which we 
must reduce from $O^*(2^n)$ to $O^*(2^{2n/3})$ to arrive at a design 
that can be placed in the Camelot framework. In the process we rely on 
a tripartite decomposition of Williams (originally presented in the 
context of 2-CSPs~\cite{Williams2005}) to decompose an edge-enumerator 
function, which forces a dependence on fast matrix multiplication in 
the per-node running times. The best known sequential algorithm runs 
in $O^*(2^n)$ time and space~\cite{BHKK2008}. Thus, up to polynomial 
factors in $n$, the design in Theorem~\ref{thm:tutte-polynomial} 
attains the optimal tradeoff $T=KE$ only if $\omega=2$. 
Furthermore, the design enables parallelization up to $K$ 
nodes for $K\leq T^{1/3}$ only.%

\subsection{Overview of techniques, earlier work, and further results}

The present Camelot framework can be seen as a continuation of 
two recent works aimed at understanding the fine-grained%
\footnote{{\em Fine-grained} in the sense that the objective is 
to obtain more precise {\em quantitative} bounds on resource 
usage compared with the more traditional coarse-grained and 
qualitative division between, e.g., polynomial time and exponential 
time in the context of classical Merlin--Arthur proofs \cite{Babai1988}.}  
complexity of 
computational problems and proof systems, initiated in a 
nondeterministic setting with a deterministic proof verifier 
by Carmosino {\em et al.}~\cite{Carmosino2016} and with the subsequent
breakthrough by Williams~\cite{Williams2016} establishing 
nontrivial qualitative bounds for a {\em randomized verifier}.
From this perspective, the Camelot framework pursues an even more
fine-grained goal by focusing also on the effort of the {\em prover}, 
with the objective of relating the 
{\em total effort to prepare the proof} to the best known 
sequential algorithms for {\em solving}%
\footnote{Without the burden of preparing a proof of correctness.}
 the problem at hand.

To design a Camelot algorithm, all it takes is to come up with 
the proof polynomial $P$ and a fast evaluation algorithm for $P$.
Our main results outlined in the previous section demonstrate
designs that attain either optimal or near-optimal speedup 
compared with the best known sequential algorithms, where
the sequential algorithms themselves are rather technical.
 
Many further designs are possible, and we find it convenient to 
introduce the techniques underlying our highlight results gradually, 
starting from more accessible designs, and with a further motivation
to illustrate the generality of the Camelot framework. 

Perhaps the more accessible Camelot designs can essentially 
be read off from the Merlin--Arthur proofs of batch evaluation 
of Williams~\cite{Williams2016}. We condense three such designs 
into the following theorem. In each case the 
problem is \#P-hard~\cite{Valiant1979}, and no 
algorithm with running time $O^*(c^n)$ for $c<2$ is known;  
algorithms with sub-exponential speedup over $O(2^n)$ are known, 
however \cite{Abboud2015,Bjorklund2012}.

\begin{Thm}
\label{thm:cnf-per-hp}
There exist Camelot algorithms that construct a proof of
size $O^*(2^{n/2})$ in time $O^*(2^{n/2})$ for each of the
following:
\begin{enumerate}
\item
The number of solutions of a CNF formula on $n$ variables.
\item
The permanent of an $n\times n$ integer matrix.
\item
The number of Hamilton cycles in an $n$-vertex graph.
\end{enumerate}
\end{Thm}
\smallskip
\noindent
(We prove Theorem~\ref{thm:cnf-per-hp} in Appendix~\ref{sect:pre-existing}.)
\smallskip

The proof polynomials $P$ associated with Theorem~\ref{thm:cnf-per-hp}
follow a design pattern where (i) the desired counts can be computed in 
essentially linear time from the available evaluations of $P$ 
(assuming the evaluations are correct!), and (ii) the polynomial $P$ is 
a composition of a vector of univariate interpolating polynomials and 
a multivariate polynomial.

Armed with this insight from batch evaluation, further
polynomials and algorithms can be extracted from earlier work on
algebraic algorithms, such as the partitioning and covering framework
of Bj\"orklund {\em et al.}~\cite{BHK2009}. The following result 
with an optimal tradeoff can be read off from the inclusion--exclusion 
formulas in this framework.

\begin{Thm}
\label{thm:set-covers}
The number of $t$-element set covers constructible from a given
family of $O^*(1)$ subsets of an $n$-element ground set can be 
computed with a Camelot algorithm that constructs a proof of
size $O^*(2^{n/2})$ in time $O^*(2^{n/2})$.
\end{Thm}
\smallskip
\noindent
(We prove Theorem~\ref{thm:set-covers} in Appendix~\ref{sect:pre-existing}.)
\smallskip

Our Camelot algorithms for the chromatic polynomial
(Theorem~\ref{thm:chromatic-polynomial}) 
and the Tutte polynomial (Theorem~\ref{thm:tutte-polynomial})
require a more subtle construction
for the proof polynomial $P$ to enable fast distributed evaluation 
at the nodes. 
In essence this is because the evaluation algorithm at each
node must work with a structured set family that may have 
size up to $2^n$ (for example, the family of independent 
sets of a graph), requiring that we rely on the structure in 
the family to arrive at $O^*(2^{n/2})$ evaluation time.
Here our design is guided by the parallel algorithm of 
Bj\"orklund {\em et al.}~\cite{BHKK2011} for computing the chromatic 
polynomial. Also the structure of the proof polynomial $P$ will be rather 
different from the polynomials underlying Theorem~\ref{thm:cnf-per-hp}
in two respects: (a) the property that we want to compute is one 
{\em coefficient} of $P$, and (b) to obtain a compact univariate proof,
we will rely on Kronecker substitution and weight tracking. 

To present the design underlying 
Theorems~\ref{thm:chromatic-polynomial}~and~\ref{thm:tutte-polynomial} in 
more technically accessible parts, as a warmup it will be convenient to 
first establish an exact-covering version of 
Theorem~\ref{thm:set-covers} that takes in an {\em exponential-sized} 
family of subsets. 

\begin{Thm}
\label{thm:set-partitions}
The number of $t$-element set partitions constructible from a given
family of $O^*(2^{n/2})$ subsets of an $n$-element ground set can be 
computed with a Camelot algorithm that constructs a proof of
size $O^*(2^{n/2})$ in time $O^*(2^{n/2})$.
\end{Thm}

The proofs of 
Theorems~\ref{thm:set-partitions},%
~\ref{thm:chromatic-polynomial},%
~and~\ref{thm:tutte-polynomial}
all rely on the same template for the proof polynomial $P$ 
which we introduce in \S\ref{sect:template},
followed by the proofs in \S\ref{sect:set-covers},
\S\ref{sect:chromatic}, and \S\ref{sect:tutte}, respectively.

Let us now move from \#P-hard problems to problems in polynomial time. 
Again it is convenient to start with a number of more accessible designs
followed by our main result. The first two designs can be read off
the batch evaluation results of Williams~\cite{Williams2016}; 
the third design is a minor
extension of the same ideas. In each case no sequential algorithm with
running time $O(n^{2-\epsilon}t^c)$ for positive constants $\epsilon$ 
and $c$ is known; sub-polynomial speedups in $n$ are 
known~\cite{Abboud2015,Alman2015,Gronlund2014,Patrascu2010}.

\begin{Thm}
\label{thm:ortho-hamming-conv3sum}
There exist Camelot algorithms that construct a proof of
size $\tilde O(nt^c)$ in time $\tilde O(nt^c)$ for each of the following:
\begin{enumerate}
\item
The number of pairs of orthogonal vectors in a given 
set of $n$ Boolean vectors of dimension $t$. (With $c=1$.)
\item
The distribution of Hamming distances between two given sets 
of $n$ Boolean vectors of dimension $t$. (With $c=2$.)
\item
The number of solutions of an instance of Convolution3SUM
consisting of $n$ integers with bit length $t$. (With $c=2$.)
\end{enumerate}
\end{Thm}
\smallskip
\noindent
(We prove Theorem~\ref{thm:ortho-hamming-conv3sum} 
in Appendix~\ref{sect:pre-existing}.)%
\smallskip

The proof polynomials underlying Theorem~\ref{thm:ortho-hamming-conv3sum}
again consist of interpolation polynomials composed with a multivariate
polynomial designed to indicate combinations of interest, where some care
is required to control the constants $c$. 
Designs analogous
to Theorem~\ref{thm:ortho-hamming-conv3sum} lead to Camelot algorithms
for counting $k$-cliques in an $n$-vertex graph that have proof 
size $\tilde O(n^{k/2})$ and run in time $n^{k/2+O(1)}$. 
(See Williams~\cite{Williams2016}.)

In contrast, our design for Theorem~\ref{thm:k-clique-camelot} relies on 
a novel arithmetic circuit design for counting 6-cliques that nests 
{\em four} levels of fast matrix multiplication so that the products at 
the top-level multiplication can (a) be evaluated in parallel {\em and} 
(b) extended to a proof polynomial $P$ by careful use of 
Yates's algorithm~\cite{Yates1937} and fast matrix multiplication to 
enable efficient evaluation. 
The top-level multiplications rely on a {\em trilinear} representation
of the matrix multiplication tensor $\bra n,n,n\ket$ to assemble the 
monomials of a $\binom{6}{2}$-linear form out of a 3-linear base and 
three 4-linear parts, whose consistency is obtained via the trilinear
representation. In essence, we show that 6-cliques can be counted
with an arithmetic circuit of essentially the {\em same size} 
but with considerably more {\em depth} than the 
Ne{\v{s}}et{\v{r}}il--Poljak~\cite{Nesetril1985} circuit. 

We present the proof of Theorem~\ref{thm:k-clique-camelot} in 
two parts. First, in \S\ref{sect:binom-6-2-form} we present the 
novel arithmetic circuits for the $\binom{6}{2}$-linear form. Then, 
in \S\ref{sect:six-cliques} we first prove Theorem~\ref{thm:k-clique}
as a warmup and then extend the top-level products in the circuit 
design to a proof polynomial $P$ and its fast evaluation algorithm.

The arithmetic circuits and the proof polynomial for the 
$\binom{6}{2}$-linear form admit also further applications beyond 
counting subgraphs, including further applications with \#P-hard 
problems. For example, as a corollary of \S\ref{sect:binom-6-2-form}
and \S\ref{sect:six-cliques} we obtain an optimal Camelot algorithm 
for enumerating variable assignments by the number of satisfied 
constraints in an instance of 2-constraint satisfaction. The
trivial sequential algorithm runs in time $O^*(\sigma^n)$ 
for $n$ variables that take values over an alphabet of $\sigma$ 
symbols, and the best known sequential algorithm runs in time 
$O^*(\sigma^{(\omega+\epsilon)n/3})$ for any constant
$\epsilon>0$~\cite{Williams2005}.

\begin{Thm}
\label{thm:enum-2csp}
For any constant $\epsilon>0$
there exists a Camelot algorithm that constructs a proof of
size $O^*(\sigma^{(\omega+\epsilon)n/6})$ in 
time $O^*(\sigma^{(\omega+\epsilon)n/6})$ 
for the enumeration of variable assignments by the number of 
satisfied constraints for a given system of 2-constraints over 
$n$ variables, with each variable taking values over an alphabet 
of size $\sigma$.
\end{Thm}
\noindent
(We prove Theorem~\ref{thm:enum-2csp} in 
Appendix~\ref{sect:further-appl}.)

\medskip
\noindent
{\em Remark.} Theorem~\ref{thm:enum-2csp} admits a generalization
to weighted instances where each $2$-constraint has a nonnegative
integer weight at most $W$. In this case both the proof size
and the per-node running time get multiplied by $W$.

Let us now discuss the techniques 
for working with sparsity and fast matrix multiplication
underlying Theorem~\ref{thm:triangle-camelot} and
Theorem~\ref{thm:triangle}. As with 
Theorem~\ref{thm:k-clique-camelot} and Theorem~\ref{thm:k-clique},
the technical contribution is to split a trilinear form of
matrix multiplication into independent parts that can be executed
in parallel, so that each part takes advantage of the sparse input.
Here the key technical tool is a novel split/sparse variant of
Yates's algorithm that we will present in 
\S\ref{sect:yates-split-sparse}. This algorithm applied to the
Itai--Rodeh design~\cite{Itai1978} gives Theorem~\ref{thm:triangle}
as an essentially immediate corollary. Furthermore, the split/sparse
algorithm admits a {\em polynomial extension} (presented 
in \S\ref{sect:yates-poly}) which in turn enables the proof
polynomial and the Camelot algorithm in 
Theorem~\ref{thm:triangle-camelot}. 

Let us conclude this section by observing that all the sequential
algorithms that we use as baseline designs to our Camelot designs
are deterministic algorithms. However, the Camelot framework extends 
in a natural way to randomized algorithms and proving the correctness
of a randomized computation if we assume the nodes have access to 
a public random string. In particular, the computation for any 
outcome of the random string is deterministic and hence verifiable
in the deterministic framework.

\section{Preliminaries}

\label{sect:preliminaries}

This section recalls notation and background results used in our main
development.

\subsection{Notation and conventions}
For a logical proposition $\Psi$ we use Iverson's bracket notation
$[\Psi]$ to indicate a 1 if $\Psi$ is true and a 0 if $\Psi$ is false.
At the risk of slight ambiguity in notation, for a nonnegative 
integer $n$ we will, however, write $[n]$ to indicate the set
$\{1,2,\ldots,n\}$.
The notation $\tilde O(\cdot)$ hides a factor polylogarithmic 
in the input size. The notation $O^*(\cdot)$ hides a factor 
polynomial in the input size.
The product over an empty set evaluates to $1$.

\subsection{Fast polynomial arithmetic}
\label{sect:fast-poly}

Let us recall the fast arithmetic toolbox 
(see von zur Gathen and Gerhard~\cite{vzGG2013}) 
for polynomials with coefficients in a finite field $\F_q$.
Computational complexity is measured in the number of arithmetic 
operations in $\F_q$.

Multiplication and division of two polynomials of degree at
most $d$ can be computed in $O(d\log d\log\log d)$ operations.
The greatest common divisor of two polynomials of degree at
most $d$ can be computed in $O(d\log^2 d\log\log d)$ operations.%
\footnote{At the same cost one can obtain any single remainder
in the sequence of remainders of the extended Euclidean algorithm,
together with the multiplier polynomials that produce the remainder.
This is serendipitous for the Gao~\cite{Gao2003} decoder which we 
review in \S\ref{sect:encode-decode}.}
Given a polynomial 
\[
P(x)=p_0+p_1x+p_2x^2\ldots+p_{d}x^{d}\in\F_q[x] 
\]
and 
points $x_0,x_1,\ldots,x_d\in\F_q$ as input, the {\em evaluation map}
\[
(P(x_0),P(x_1),\ldots,P(x_d))\in\F_q^{d+1}
\]
can be computed in $O(d\log^2 d\log\log d)$ operations. Dually, given 
distinct points $x_0,x_1,\ldots,x_d\in\F_q$ and values 
$y_0,y_1,\ldots,y_d\in\F_q$ as input, the {\em interpolation map}
computes the coefficients $(p_0,p_1,\ldots,p_d)\in\F_q^{d+1}$ 
of a polynomial $P(x)$ such that $P(x_i)=y_i$ for 
all $i=0,1,\ldots,d$ in $O(d\log^2 d\log\log d)$ operations. 

\subsection{Reed--Solomon codes, fast encoding and decoding}

\label{sect:encode-decode}

We work with univariate polynomial codes in nonsystematic form 
as originally discovered by Reed and Solomon~\cite{Reed1960} to allow for
fast decoding based on polynomial arithmetic (cf.~Gao~\cite{Gao2003}).

Fix a finite field $\F_q$ and any $1\leq e\leq q$ distinct elements 
$x_1,x_2,\ldots,x_e\in\F_q$. The parameter $e$ is the {\em length} of 
the code. To {\em encode} a {\em message} $(p_0,p_1,\ldots,p_d)\in\F_q^{d+1}$ 
consisting of $d+1$ {\em symbols}, $1\leq d+1\leq e$, 
form the message polynomial $P(x)=p_0+p_1x+p_2x^2+\ldots+p_dx^d$ and 
the {\em codeword}
\[
(P(x_1),P(x_2),\ldots,P(x_e))\in\F_q^e\,.
\]
The codeword can be computed from the message in 
$O(e\log^2 e\log\log e)$ operations using fast interpolation
(see \S\ref{sect:fast-poly}).

Since a nonzero polynomial of degree $d$ has at most $d$ roots, 
any two codewords either agree in at most $d$ symbols or are identical. 
Thus, we can uniquely {\em decode} the message from any 
{\em received} word $(r_1,r_2,\ldots,r_e)\in\F_q^n$ that 
differs from the codeword in at most $(e-d-1)/2$ symbols.

To decode the received word (or assert decoding failure), we use the 
following algorithm of Gao~\cite{Gao2003}.
As a precomputation step, compute the polynomial 
$G_0(x)=\prod_{i=1}^e(x-x_i)\in\F_q[x]$.%
\footnote{When $e=q$, we have $G_0(x)=x^q-x$ which suffices for 
our applications in this paper.}{}
To decode $(r_1,r_2,\ldots,r_e)\in\F_q^e$, 
first interpolate the unique polynomial $G_1(x)\in\F_q[x]$ 
of degree at most $e-1$ with $G_1(x_i)=r_i$ for all $i=1,2,\ldots,e$.
Second, apply the extended Euclidean algorithm to the polynomials
$G_0(x)$ and $G_1(x)$, but stop as soon as the remainder $G(x)$ 
has degree less than $(e+d+1)/2$.
At this point suppose that
\[
U(x)G_0(x)+V(x)G_1(x)=G(x)\,.
\]
Finally, divide $G(x)$ by $V(x)$ to obtain 
\[
G(x)=P(x)V(x)+R(x)\,,
\]
where the degree of $R(x)$ is less than the degree of $V(x)$. 
If $R(x)=0$ and $P(x)$ has degree at most $d$, output $P(x)$ as the
result of decoding; otherwise assert decoding failure. 
We observe that decoding runs in $O(e\log^2 e\log\log e)$ operations
using fast polynomial arithmetic.

\section{A split/sparse variant of Yates's algorithm}

This section presents a variant of Yates's algorithm~\cite{Yates1937}
that accepts {\em sparse} input and {\em splits} its output into 
multiple parts that can be produced independently of each other,
in parallel. This variant can essentially
be read off the split/sparse fast zeta transform of 
Bj\"orklund~{\em et~al.}~\cite{BHKK2011}.
Furthermore, the algorithm readily admits a polynomial extension 
that will prove serendipitous for the design of Camelot algorithms
(see \S\ref{sect:triangle}).

\subsection{Yates's algorithm}

\label{sect:yates-classical}

Let us start by recalling the classical Yates's algorithm. 
For positive integers $s,t,k$, 
Yates's algorithm~\cite{Yates1937} multiplies a 
given $s^k\times 1$ vector $x$ with a structured $t^k\times s^k$ 
matrix $M$ to produce as output a $t^k\times 1$ vector $y=Mx$. 
The structural assumption about the matrix $M$ is that it is 
a Kronecker power $M=A^{\otimes k}$ of a small $t\times s$ 
matrix $A$ with entries $\alpha_{ij}$ for $i=1,2,\ldots,t$
and $j=1,2,\ldots,s$. All arithmetic takes place in a field $\F$.

For positive integers $d,k$ we shall tacitly identify 
each $j\in [d^k]$ with $(j_1,j_2,\ldots,j_k)\in [d]^k$.
Indeed, we can view $j$ as a $k$-digit integer in base $d$, 
where $j_1,j_2,\ldots,j_k$ are the $k$ digits.

Given $x$ and $A$ as input, Yates's algorithm 
seeks to compute, for all $i\in[t^k]$, the value
\begin{equation}
\label{eq:yates-output}
y_i=\sum_{j\in[s^k]} \alpha_{i_1j_1}\alpha_{i_2j_2}\cdots \alpha_{i_kj_k}x_j\,.
\end{equation}
The algorithm proceeds in $k$ levels, where level 
$\ell=1,2,\ldots,k$ takes as input a $t^{\ell-1}s^{k+1-\ell}\times 1$ vector $y^{(\ell-1)}$
and produces as output a $t^{\ell}s^{k-\ell}\times 1$ 
vector $y^{(\ell)}$.
The input to the first level is $y^{(0)}=x$. 
The vector $y^{(\ell)}$ is defined for all 
$i_1,i_2,\ldots,i_{\ell}\in [t]$ 
and
$j_{\ell+1}j_{\ell+2}\cdots j_k\in [s]$ 
by
\begin{equation}
\label{eq:yates-intermediate}
y^{(\ell)}_{i_1i_2\cdots i_\ell j_{\ell+1}j_{\ell+2}\cdots j_k}
=
\sum_{j_1,j_2,\ldots,j_\ell\in[s]}
\alpha_{i_1j_1}\alpha_{i_2j_2}\cdots \alpha_{i_\ell j_\ell}x_j\,.
\end{equation}
It is immediate that $y^{(k)}=y$ as desired.
To compute $y^{(\ell)}$ given $y^{(\ell-1)}$ as input, we 
observe that for all
$i_1,i_2,\ldots,i_{\ell}\in [t]$ 
and
$j_{\ell+1}j_{\ell+2}\cdots j_k\in [s]$ 
it holds that
\begin{equation}
\label{eq:yates-recurrence}
y^{(\ell)}_{i_1i_2\cdots i_\ell j_{\ell+1}j_{\ell+2}\cdots j_k}
=
\sum_{j_\ell\in[s]}
\alpha_{i_\ell j_\ell}
y^{(\ell-1)}_{i_1i_2\cdots i_{\ell-1} j_{\ell}j_{\ell+1}\cdots j_k}\,.
\end{equation}
In essence, \eqref{eq:yates-recurrence} proceeds from the input
$x=y^{(0)}$ towards the output $y=y^{(k)}$ by evaluating
``one of the $k$ nested sums at a time'', as is readily verified 
using induction on $\ell$ and \eqref{eq:yates-intermediate}.

From \eqref{eq:yates-intermediate} it follows that
Yates's algorithm runs in $O((s^{k+1}+t^{k+1})k)$ operations and 
working space for $O(s^k+t^k)$ field elements.

\subsection{The split/sparse algorithm}

\label{sect:yates-split-sparse}

Let us assume that $t\geq s$. 
Suppose the input vector $x$ is {\em sparse} so that $x_j\neq 0$ only 
if $j\in D$ for a nonempty set $D\subseteq [s^k]$. Accordingly, we
assume that the input consists of $O(|D|)$ field elements.
We seek to produce the $t^k$ elements of the output vector $y$ in 
roughly $t^k/|D|$ independent parts so that each part consists 
of roughly $|D|$ entries of $y$. In more precise terms, 
the number of entries of $y$ in each part is the 
minimum multiple of $t$ at least $|D|$.

The algorithm in particular {\em splits} the output indices
$i_1i_2\cdots i_k$ into two groups, $i_1i_2\cdots i_\ell$ 
and $i_{\ell+1}i_{\ell+2}\cdots i_k$.
The outer iteration works with the latter index group whereas 
the inner iteration works with the former group and executes 
the classical Yates's algorithm. 
The pseudocode is as follows:
\begin{enumerate}
\item
For each $i_{\ell+1},i_{\ell+2},\ldots,i_k\in [t]$ do the following:
\begin{enumerate}
\item
For each $j_1,j_2,\ldots,j_\ell\in [s]$,
set
\[
x^{(\ell)}_{j_1j_2\cdots j_\ell}
\leftarrow
0\,.
\]
\item
For each $j\in D$, set 
\[
x^{(\ell)}_{j_1j_2\cdots j_\ell}
\leftarrow
x^{(\ell)}_{j_1j_2\cdots j_\ell}
+
\alpha_{i_{\ell+1}j_{\ell+1}}
\alpha_{i_{\ell+2}j_{\ell+2}}
\cdots
\alpha_{i_{k}j_{k}}
x_j\,.
\]
\item
Use (classical) Yates's algorithm 
to compute $u^{(\ell)}\leftarrow A^{\otimes\ell} x^{(\ell)}$.
\item
For each $i_1,i_2,\ldots,i_\ell\in [t]$, 
output $u^{(\ell)}_{i_1i_2\cdots i_\ell}$ as the value
$y_i=y_{i_1i_2\cdots i_k}$.
\end{enumerate}
\end{enumerate}
Now take $\ell=\lceil\log_t |D|\rceil$ to achieve the desired
output in parts.

\medskip
\noindent 
{\em Resource consumption.}
First, observe that the $t^{k-\ell}$ iterations of the outer loop are
independent of each other and can be executed in parallel.
Each iteration of the outer loop requires 
$O((t^{\ell+1}+s^{\ell+1})\ell+|D|)$ operations and working space for 
$O(t^\ell+s^\ell+|D|)$ field elements. 

Recalling that $\ell=\lceil(\log |D|)/(\log t)\rceil$
and that $t\geq s$, we observe 
that each iteration of the outer loop takes
$O(|D|t^2(\log |D|)/(\log t))$ operations and working 
space for $O(t|D|)$ field elements.

In total the algorithm takes 
$O\bigl(t^{k+2}(\log |D|)/(\log t)\bigr)$ operations to deliver 
its output in $\Omega(t^{k-1}/|D|)$ independent parts, each 
consisting of $O(t|D|)$ field elements.

\subsection{Polynomial extension}

\label{sect:yates-poly}

Let us now develop a {\em polynomial extension} of the 
split/sparse algorithm in the previous section. 
In particular, we replace the outer loop 
with a polynomial indeterminate $z$. That is, in place of iterating
the outer loop, we perform repeated {\em evaluations} at points
$z=z_0\in\F$, with the intuition that values%
\footnote{We assume specific $t^{k-\ell}$ values in $\F$ 
have been identified with elements of $[t^{k-\ell}]$. 
For example, such an identification is immediate if
$\F$ is a finite field (of large enough) prime order.}{}
$z_0\in [t^{k-\ell}]$ deliver precisely the output that is 
produced by the outer loop of the split/sparse algorithm 
during iteration 
$(i_{\ell+1},i_{\ell+2},\ldots,i_k)\in [t^{k-\ell}]$.

Accordingly, we replace references to the matrix $A$ 
for the {\em outer} indices 
$i_{\ell+1},i_{\ell+2},\ldots,i_k\in [t]$ 
with $|D|$ polynomials $\alpha_j(z)$, one polynomial for 
each $j\in D$. 

We present the pseudocode first in abstract form (with
an indeterminate $z$), with the understanding that concrete executions
of the algorithm carry out an {\em evaluation} at a given point 
$z=z_0\in\F$. We continue to assume that $t\geq s$.
The abstract pseudocode for the polynomial extension is as follows:
\begin{enumerate}
\item
For each $j_1,j_2,\ldots,j_\ell\in [s]$,
set
\[
x^{(\ell)}_{j_1j_2\cdots j_\ell}(z)
\leftarrow
0\,.
\]
\item
For each $j\in D$, set 
\[
x^{(\ell)}_{j_1j_2\cdots j_\ell}(z)
\leftarrow
x^{(\ell)}_{j_1j_2\cdots j_\ell}(z)
+
\alpha_j(z)
x_j\,.
\]
\item
Use (classical) Yates's algorithm 
to compute $u^{(\ell)}(z)\leftarrow A^{\otimes\ell}x^{(\ell)}(z)$.
\item
For each $i_1,i_2,\ldots,i_\ell\in [t]$, 
output $u^{(\ell)}_{i_1i_2\cdots i_\ell}(z)$.
\end{enumerate}
We observe that $u^{(\ell)}_{i_1i_2\ldots i_\ell}(z)$
is a polynomial whose degree is at most the maximum of the
degrees of the polynomials $\alpha_j(z)$ for $j\in D$. 
Let us proceed to define these polynomials.

Towards this end, for $i\in [t^{k-\ell}]$, let us write 
\begin{equation}
\label{eq:yates-lag-R}
\Phi_i(z)=\prod_{\substack{w=1\\w\neq i}}^{t^{k-\ell}} \frac{z-w}{i-w}
\end{equation}
for the Lagrange interpolation polynomial of degree at most 
$t^{k-\ell}-1$ that is $1$ at $i$ and $0$ elsewhere in $[t^{k-\ell}]$. 
For each $j\in D$ the univariate polynomial
\begin{equation}
\label{eq:yates-coeff-poly}
\alpha_{j}(z)=\sum_{i\in[t^{k-\ell}]} \alpha_{i_{\ell+1}j_{\ell+1}}\alpha_{i_{\ell+2}j_{\ell+2}}\cdots \alpha_{i_kj_k}\Phi_i(z)
\end{equation}
now interpolate over the desired coefficients.
Indeed, for all $i\in [t^{k-\ell}]$ and $j\in D$ we observe that
\[
\alpha_{j}(i)=\sum_{i\in[t^{k-\ell}]} \alpha_{i_{\ell+1}j_{\ell+1}}\alpha_{i_{\ell+2}j_{\ell+2}}\cdots \alpha_{i_kj_k}\,.
\]

To turn the abstract pseudocode into a concrete evaluation algorithm,
we need a fast algorithm for evaluating the $|D|$ polynomials 
$\alpha_j(z)$ at a specific point $z=z_0$. 
Suppose we are given $z_0\in\F$ as input. We need 
to compute $\alpha_j(z_0)\in\F$ for each $j\in D$. Recalling 
\eqref{eq:yates-coeff-poly}, we have, for all $j\in D$,
\begin{equation}
\label{eq:yates-kron-vec}
\alpha_{j}(z_0)=\sum_{i\in[t^{k-\ell}]} \alpha_{i_{\ell+1}j_{\ell+1}}\alpha_{i_{\ell+2}j_{\ell+2}}\cdots \alpha_{i_kj_k}\Phi_i(z_0)\,.
\end{equation}

We observe that \eqref{eq:yates-kron-vec} in fact
describes a matrix-vector multiplication:
we multiply the $|D|\times t^{k-\ell}$ matrix with entries 
$\alpha_{i_{\ell+1}j_{\ell+1}}\alpha_{i_{\ell+2}j_{\ell+2}}\cdots \alpha_{i_kj_k}$
with the $t^{k-\ell}\times 1$ vector with entries $\Phi_i(z_0)$
to obtain as a result the $|D|\times 1$ vector with 
entries $\alpha_j(z_0)$.

Let us now observe that the matrix entries 
$\alpha_{i_{\ell+1}j_{\ell+1}}\alpha_{i_{\ell+2}j_{\ell+2}}\cdots \alpha_{i_kj_k}$ only require the last $k-\ell$ components of
each index $j\in D$. Accordingly, it suffices to consider
a $s^{k-\ell}\times t^{k-\ell}$ matrix and then use 
$j\in D$ to {\em index} the computed values
to obtain $\alpha_j(z_0)$ for each $j\in D$.
In particular, we can use classical Yates's algorithm to compute
the value $\alpha_{j_{\ell+1}j_{\ell+2}\cdots j_k}(z_0)$ for each
$(j_{\ell+1},j_{\ell+2},\cdots,j_k)\in[s]^{k-\ell}$.
Recalling that we assume $t\geq s$, 
this matrix--vector multiplication takes 
$O(t^{k-\ell+1}(k-\ell)+|D|)$ operations and 
working space for $O(t^{k-\ell}+|D|)$ field elements. 

To initialize Yates's algorithm, we still require the 
$t^{k-\ell}\times 1$ vector, so let us describe how to compute 
the values $\Phi_i(z_0)$ for $i\in[t^{k-\ell}]$. 
If $z_0\in[t^{k-\ell}]$, the computation is immediate 
(insert a 1 to position $z_0$ and fill the rest of the 
vector with 0s), so let us assume that $z_0\notin[t^{k-\ell}]$. 
First, precompute the values 
$F_0,F_1,\ldots,F_{t^{k-\ell}-1}$ with the recurrence
\[
F_0=1,\quad F_\ell=\ell\cdot F_{\ell-1}\,.
\]
Second, precompute the product
\[
\Gamma(z_0)=\prod_{\ell=1}^{t^{k-\ell}}(z_0-\ell)\,.
\]
Finally, observe from \eqref{eq:yates-lag-R} that we can compute,
for each $i\in[t^{k-\ell}]$, 
\[
\Phi_i(z_0)=\frac{1}{(-1)^{t^{k-\ell}-i}F_{i-1}F_{t^{k-\ell}-i}}\cdot
               \frac{\Gamma(z_0)}{(z_0-i)}\,.
\]
This initialization takes in total $O(t^{k-\ell})$ operations and 
space.

For a $z_0\in\F$, a vector $x$, and the matrix $A$ given as input, 
the polynomial extension algorithm thus runs in 
$O(|D|t^2(\log |D|)/(\log d)+t^{k-\ell+1}(k-\ell))$ operations and 
working space for $O(t|D|+t^{k-\ell})$ field elements.
The algorithm produces as output the values
$u^{(\ell)}_{i_1i_2\cdots i_\ell}(z_0)$
for each $i_1,i_2,\ldots,i_\ell\in [t]$. 

Viewed as a polynomial in $z$, each polynomial
$u^{(\ell)}_{i_1i_2\cdots i_\ell}(z)$
for $i_1,i_2,\ldots,i_\ell\in [t]$
has degree at most $t^{k-\ell}-1$.
Furtermore, for all $i\in[t^k]$ we have
$y_i=u^{(\ell)}_{i_1i_2\cdots i_\ell}(i_{\ell+1}i_{\ell+1}\cdots i_k)$.
Here $y=A^{\otimes k}x$.
That is, the polynomial extension algorithm recovers
the output of the split/sparse algorithm.

\section{Fast and space-efficient evaluation of the $\binom{6}{2}$-linear form}

\label{sect:binom-6-2-form}

This section studies a multilinear form that integrates over 
a function or constraint system with only pairwise interactions
between variables by decomposing the interactions into $\binom{6}{2}=15$ 
parts. The advantage of such a decomposition is that one can use fast 
matrix multiplication to arrive at a nontrivial arithmetic circuit 
for fast integration. An example application is counting small subgraphs
of a graph, which we will illustrate in more detail in the next section.

\subsection{The problem and the Ne{\v{s}}et{\v{r}}il--Poljak formula}

Let $\chi$ be an%
\footnote{The present formulation admits an immediate generalization
to $\binom{6}{2}$ distinct $N\times N$ matrices, but to keep
the notation concise, we work with a single matrix $\chi$.}{}
$N\times N$ matrix over a commutative ring $\R$.
We seek to compute the $\binom{6}{2}$-linear form
\vspace*{-0.5mm}
\begin{equation}
\label{eq:six-sum}
X_{\binom{6}{2}}=\sum_{a,b,c,d,e,f}
\chi_{ab}\chi_{ac}\chi_{ad}\chi_{ae}\chi_{a\!f}
\chi_{bc}\chi_{bd}\chi_{be}\chi_{b\!f}
\chi_{cd}\chi_{ce}\chi_{c\!f}
\chi_{de}\chi_{d\!f}
\chi_{e\!f}\,.
\end{equation}

A direct evaluation of \eqref{eq:six-sum} takes $O(N^6)$ operations
and $O(N^2)$ space.
Ne{\v{s}}et{\v{r}}il and Poljak \cite{Nesetril1985} observe that
we can precompute the three $N^2\times N^2$ matrices 
\[
\begin{split}
U_{ab,cd}&=\chi_{ab}\chi_{ac}\chi_{ad}\chi_{bc}\chi_{bd}\,,\\
S_{ab,e\!f}&=\chi_{ae}\chi_{a\!f}\chi_{be}\chi_{b\!f}\chi_{e\!f}\,,\\
T_{cd,e\!f}&=\chi_{cd}\chi_{ce}\chi_{c\!f}\chi_{de}\chi_{d\!f}\,,
\end{split}
\]
and then use fast matrix multiplication to compute 
\[
X_{\binom{6}{2}}=\sum_{a,b,c,d}U_{ab,cd}V_{ab,cd},
\qquad
V_{ab,cd}=\sum_{e,f} S_{ab,e\!f}T_{cd,e\!f}\,.
\]
This takes $O(N^{2\omega+\epsilon})$ arithmetic operations 
and $O(N^4)$ space for any constant $\epsilon>0$.

\subsection{A new summation formula and its complexity}
\label{sect:sum-formula}

Our main contribution in this section is a new
circuit design that matches the Ne{\v{s}}et{\v{r}}il--Poljak design
in the asymptotic number of arithmetic operations
but reduces the space complexity from $O(N^4)$ to $O(N^2)$. 
In particular, the space complexity is linear in the input size. 
Furthermore, the new design is easily parallelizable and extendable
to a proof polynomial, as we will witness in the next section.

The new design works with the following explicit decomposition of 
the matrix multiplication tensor. For $r=1,2,\ldots,R$,
let $\alpha_{de}(r),\beta_{e\!f}(r),\gamma_{d\!f}(r)$ be ring
elements that satisfy the polynomial identity
\begin{equation}
\label{eq:tri-consistency}
\!\sum_{d,e,f}u_{de}v_{e\!f}w_{d\!f}\!
=
\!\sum_{r=1}^R
\biggl(
\sum_{d,e'}\alpha_{de'}(r)u_{de'}
\!\biggr)
\biggl(
\sum_{e,f'}\beta_{e\!f'}(r)v_{e\!f'}
\!\biggr)
\biggl(
\sum_{d',f}\gamma_{d'\!f}(r)w_{d'\!f}
\!\biggr)\,.\!\!\!\!\!
\end{equation}
We can assume that $R=O(N^{\omega+\epsilon/2})$ for an arbitrary constant
$\epsilon>0$, where $\omega$ is the limiting exponent for the tensor rank 
of the matrix multiplication tensor $\bra n,n,n\ket$ in 
$\R$~\cite{Burgisser1997,Landsberg2012}.%

The new design is as follows.
For each $r=1,2,\ldots,R$, compute, using fast matrix multiplication,
\begin{align}
\notag
A_{ab}(r)&=\sum_{d}\chi_{ad}\chi_{bd}H_{ad}(r)
\,,&
H_{ad}(r)&=\sum_{e'}\alpha_{de'}(r)\chi_{ae'}\chi_{de'}
\,,\\
\label{eq:abc}
B_{bc}(r)&=\sum_{e}\chi_{be}\chi_{ce}K_{be}(r)
\,,&
K_{be}(r)&=\sum_{f'}\beta_{e\!f'}(r)\chi_{b\!f'}\chi_{e\!f'}
\,,\\
\notag
C_{ac}(r)&=\sum_{f}\chi_{a\!f}\chi_{c\!f}L_{c\!f}(r)
\,,&
L_{c\!f}(r)&=\sum_{d'}\gamma_{d'\!f}(r)\chi_{cd'}\chi_{d'\!f}
\,.
\end{align}
Finally, compute, again using fast matrix multiplication,
\begin{align}
\label{eq:p-formula}
P(r)&=\sum_{a,b}\chi_{ab}A_{ab}(r)Q_{ab}(r)\,,&
Q_{ab}(r)&=\sum_{c}\chi_{ac}\chi_{bc}B_{bc}(r)C_{ac}(r)\,.
\end{align}
Each term $P(r)$ takes $O(N^{\omega+\epsilon/2})$ operations and $O(N^2)$ 
space to compute. 

\begin{Thm}
\label{thm:main}
$X_{\binom{6}{2}}=\sum_{r=1}^R P(r)$.
\end{Thm}

\begin{Proof}
Expanding and applying \eqref{eq:tri-consistency} for each $a,b,c$ in turn, 
we have
\[
\begin{split}
\sum_{r=1}^R P(r)&=
\sum_{r=1}^R
\sum_{a,b,c}
\chi_{ab}\chi_{ac}\chi_{bc}
A_{ab}(r)B_{bc}(r)C_{ac}(r)
\\
&=
\sum_{r=1}^R
\sum_{a,b,c}
\chi_{ab}\chi_{ac}\chi_{bc}
\sum_{d,e'}\alpha_{de'}(r)\chi_{ad}\chi_{ae'}\chi_{bd}\chi_{de'}\\
&\qquad\qquad\qquad\qquad\quad
\sum_{e,f'}\beta_{e\!f'}(r)\chi_{be}\chi_{b\!f'}\chi_{ce}\chi_{e\!f'}\\
&\qquad\qquad\qquad\qquad\quad
\sum_{d',f,}\gamma_{d'\!f}(r)\chi_{a\!f}\chi_{cd'}\chi_{c\!f}\chi_{d'\!f}\\
&=
\sum_{a,b,c}
\chi_{ab}\chi_{ac}\chi_{bc}
\sum_{r=1}^R
\biggl(
\sum_{d,e'}
\alpha_{de'}(r)\chi_{ad}\chi_{ae'}\chi_{bd}\chi_{de'}
\biggr)\\
&\qquad\qquad\qquad\qquad\quad
\biggl(
\sum_{e,f'}
\beta_{e\!f'}(r)\chi_{be}\chi_{b\!f'}\chi_{ce}\chi_{e\!f'}
\biggr)\\
&\qquad\qquad\qquad\qquad\quad
\biggl(
\sum_{d',f}
\gamma_{d'\!f}(r)\chi_{a\!f}\chi_{cd'}\chi_{c\!f}\chi_{d'\!f}
\biggr)
\\
&=
\sum_{a,b,c}
\chi_{ab}\chi_{ac}\chi_{bc}
\sum_{d,e,f}
\chi_{ad}\chi_{ae}\chi_{a\!f}
\chi_{bd}\chi_{be}\chi_{b\!f}
\chi_{cd}\chi_{ce}\chi_{c\!f}
\chi_{de}\chi_{d\!f}
\chi_{e\!f}
\,.\\[-0.8\baselineskip]
\end{split}
\]
\end{Proof}

Thus, we can compute \eqref{eq:six-sum} from $\chi$ 
in $O(N^{2\omega+\epsilon})$ operations and $O(N^2)$ space.
Furthermore, easy parallelization using up to $O(N^{\omega+\epsilon/2})$ 
compute nodes is now possible since the values $P(r)$ can be 
computed independently of each other.

\section{A Camelot algorithm for counting small cliques}

\label{sect:six-cliques}

This section proves Theorem~\ref{thm:k-clique-camelot} 
and Theorem~\ref{thm:k-clique}. In particular, we demonstrate
how to reduce counting $k$-cliques 
into an evaluation of the $\binom{6}{2}$-linear form, and 
then develop a proof polynomial $P(x)$ for the 
$\binom{6}{2}$-linear form, together with a fast evaluation 
algorithm that computes $P(x_0)\pmod q$ for given $x_0\in\Z_q$. 

\subsection{Reduction to the $\binom{6}{2}$-linear form}

Suppose that $6$ divides $k$.
We start with a routine reduction from counting $k$-cliques in 
a given $n$-vertex graph $G$ to the task of evaluating 
the $\binom{6}{2}$-linear form for a particular matrix $\chi$
derived from $G$.

Let $\chi$ be an $N\times N$ integer matrix with 
$N=\binom{n}{k/6}\leq n^{k/6}$ 
and entries defined for all $A,B\in\binom{V(G)}{k/6}$ by
\[
\chi_{AB}
=
\bigl[
\text{$A\cup B$ is a clique in $G$ and $A\cap B=\emptyset$}
\bigr]\,.
\]
That is, the entries of $\chi$ indicate all pairs of
$k/6$-cliques in $G$ that together form a $k/3$-clique.

The $\binom{6}{2}$-linear form $X_{\binom{6}{2}}$ with 
input $\chi$ now counts each $k$-clique in $G$ exactly 
$\binom{k}{k/6,\,k/6,\,k/6,\,k/6,\,k/6,\,k/6}$ times.
Recalling that $N\leq n^{k/6}$, 
Theorem~\ref{thm:k-clique} now follows by 
applying the algorithm in \S\ref{sect:sum-formula} to 
compute $X_{\binom{6}{2}}$.

\subsection{A proof polynomial for the $\binom{6}{2}$-linear form}

We now turn the counting algorithm in the previous section into 
a Camelot algorithm. First, we present a proof polynomial
$P(x)$ in the Camelot framework that enables us to compute
and verify the value $X_{\binom{6}{2}}$ from the common input $\chi$.%

We define a univariate polynomial $P(x)$ so that we can 
recover $X_{\binom{6}{2}}$ via Theorem~\ref{thm:main} using 
evaluations of $P(x)$ at integer points $x=1,2,\ldots,R$. 
We proceed by extending the coefficients 
$\alpha_{de}(r),\beta_{e\!f}(r),\gamma_{d\!f}(r)$ 
into interpolation polynomials and then extending
\eqref{eq:abc} and \eqref{eq:p-formula} into polynomials. 

Towards this end, for $r=1,2,\ldots,R$, let us write 
\begin{equation}
\label{eq:lag-R}
\Lambda_r(x)=\prod_{\substack{j=1\\j\neq r}}^R \frac{x-j}{r-j}
\end{equation}
for the Lagrange interpolation polynomials of degree at most $R$ for 
the points $1,2,\ldots,R$. The univariate polynomials
\begin{equation}
\label{eq:coeff-poly}
\begin{split}
\alpha_{de}(x)&=\sum_{r=1}^R \alpha_{de}(r)\Lambda_r(x)\,,\\
\beta_{e\!f}(x)&=\sum_{r=1}^R \beta_{e\!f}(r)\Lambda_r(x)\,,\\
\gamma_{d\!f}(x)&=\sum_{r=1}^R \gamma_{d\!f}(r)\Lambda_r(x)\,
\end{split}
\end{equation}
now interpolate over the coefficients 
$\alpha_{de}(r),\beta_{e\!f}(r),\gamma_{d\!f}(r)$ 
when $x=1,2,\ldots,R$.
From \eqref{eq:coeff-poly}, \eqref{eq:abc}, and \eqref{eq:p-formula} 
it is immediate that the univariate polynomials
\begin{align}
\notag
A_{ab}(x)&=\sum_{d}\chi_{ad}\chi_{bd}H_{ad}(x)
\,,&
H_{ad}(x)&=\sum_{e'}\alpha_{de'}(x)\chi_{ae'}\chi_{de'}
\,,\\
\label{eq:abc-poly}
B_{bc}(x)&=\sum_{e}\chi_{be}\chi_{ce}K_{be}(x)
\,,&
K_{be}(x)&=\sum_{f'}\beta_{e\!f'}(x)\chi_{b\!f'}\chi_{e\!f'}
\,,\\
\notag
C_{ac}(x)&=\sum_{f}\chi_{a\!f}\chi_{c\!f}L_{c\!f}(x)
\,,&
L_{c\!f}(x)&=\sum_{d'}\gamma_{d'\!f}(x)\chi_{cd'}\chi_{d'\!f}
\,,
\end{align}
and 
\begin{align}
\label{eq:p-formula-poly}
P(x)&=\sum_{a,b}\chi_{ab}A_{ab}(r)Q_{ab}(x)\,,&
Q_{ab}(x)&=\sum_{c}\chi_{ac}\chi_{bc}B_{bc}(x)C_{ac}(x)
\end{align}
have degree at most $3R$ and the evaluations of $P(x)$ 
at $x=1,2,\ldots,R$ satisfy Theorem~\ref{thm:main}. 
Thus, $P(x)$ is a proof polynomial that we can employ in
the Camelot framework. For the modulus $q$ we can select 
one or more primes $q\geq 3R+1$ to enable interpolation and
reconstruction of $X_{\binom{6}{2}}$ over the integers. 
In particular we can assume $q=O(R)$. Indeed, 
since $q\geq N^2$ and $\chi$ has entries in $\{0,1\}$, 
we can recover the integer $0\leq X_{\binom{6}{2}}=\sum_{r=1}^R P(r)\leq N^6$ 
from evaluations $\sum_{r=1}^R P(r)\pmod q$ for at most $O(1)$ 
distinct primes $q\geq 3R+1$ using the Chinese Remainder Theorem. 
This gives proof size $\tilde O(R)$. 

\subsection{The evaluation algorithm}

\label{sect:small-clique-eval}

To complete the Camelot algorithm it remains to describe how each node 
evaluates, modulo $q$, the proof polynomial $P(x)$ at a given point. 
Let $x_0\in\Z_q$ be given. We seek to compute $P(x_0)\pmod q$
in time $O(N^{\omega+\epsilon})$.

Let us first assume that we have available the values 
$\alpha_{de}(x_0),\beta_{e\!f}(x_0),\gamma_{d\!f}(x_0)$
for all $d,e,f$. 
(This is a nontrivial assumption that we will justify in what follows.)
Then, using fast matrix multiplication to evaluate 
\eqref{eq:abc-poly} and \eqref{eq:p-formula-poly} at $x=x_0$, 
we obtain $P(x_0)$ in $O(N^{\omega+\epsilon})$ operations 
and $O(N^2)$ space. Arithmetic operations on scalars modulo $q=O(R)$ can
be implemented in $\tilde O(1)$ time and space, which is subsumed
in the choice of $\epsilon>0$. 
Thus, we are done assuming we can justify our earlier assumption.

To compute the values 
$\alpha_{de}(x_0),\beta_{e\!f}(x_0),\gamma_{d\!f}(x_0)$ for all 
$d,e,f$, we proceed to take a look at the detailed structure of the 
coefficients $\alpha_{de}(r),\beta_{e\!f}(r),\gamma_{d\!f}(r)$.
We will focus on the coefficients
$\alpha_{de}(r)$, the cases for $\beta_{e\!f}(r)$ and $\gamma_{d\!f}(r)$
are symmetric. 

By the properties of tensor rank and the fact that matrix multiplication
tensors are closed under taking of Kronecker products, 
for any constant $\epsilon>0$ we can assume without loss 
of generality that $R=O(N^{\omega+\epsilon})$ and that 
there exist positive 
integer constants $R_0,N_0$ and a positive integer $t$ 
such that $N=N_0^t$ and $R=R_0^t$. In particular, we may identify
the indices $d,e$ and the index $r$ with $t$-digit integers 
in base $N_0$ and $R_0$, respectively, so that if we write
$d_j,e_j$ and $r_j$ for the $j$th digit of $d,e$ and $r$, respectively,
the coefficient $\alpha_{de}(r)$ has the Kronecker product form 
\begin{equation}
\label{eq:alpha-kron}
\alpha_{de}(r)=\prod_{j=1}^t\alpha^{(0)}_{d_je_j}(r_j)\,,
\end{equation}
where $\alpha^{(0)}$ is a matrix of size $N_0^2\times R_0$ with integer
entries. Thus, we may view \eqref{eq:alpha-kron} as an 
integer matrix of size $N^2\times R$ which has been obtained 
as the $t$-fold Kronecker power of the constant-sized-and-entried 
matrix $\alpha^{(0)}$. 

Let us now describe how to compute the $N^2$ coefficients 
$\alpha_{de}(x_0)\pmod q$ for all $d,e$. 
(The computations for $\beta_{e\!f}(x_0)$ and 
$\gamma_{d\!f}(x_0)$ are symmetric.)

Let us recall from 
\eqref{eq:coeff-poly} that $\alpha_{de}(x)$ is an interpolating
polynomial for the values \eqref{eq:alpha-kron}. 
That is, we have, for all $d,e$, 
\begin{equation}
\label{eq:kron-vec}
\alpha_{de}(x_0)=\sum_{r=1}^R \alpha_{de}(r)\Lambda_r(x_0)\pmod q\,.
\end{equation}
We observe that \eqref{eq:kron-vec} in fact
describes a matrix-vector multiplication:
we multiply the $N^2\times R$ matrix with entries $\alpha_{de}(r)$ 
with the $R\times 1$ vector with entries $\Lambda_r(x_0)$. 
Because of the Kronecker structure \eqref{eq:alpha-kron}, 
this matrix-vector product can be computed in 
$O(Rt)$ operations using Yates's algorithm 
(\S\ref{sect:yates-classical}).

To initialize Yates's algorithm, we require the $R\times 1$ vector, 
so let us start by computing the values $\Lambda_r(x_0)\pmod q$ 
for $r=1,2,\ldots,R$. If $x_0\in\{1,2,\ldots,R\}$ the computation is 
immediate (insert a 1 to position $x_0$ and fill the rest of the 
vector with 0s), so let us assume that $x_0\notin\{1,2,\ldots,R\}$. 
First, precompute the values $F_0,F_1,\ldots,F_{R-1}$ with the
recurrence
\[
F_0=1,\quad F_j=j\cdot F_{j-1}\pmod q\,.
\]
Second, precompute the product
\[
\Gamma(x_0)=\prod_{j=1}^R(x_0-j)\pmod q\,.
\]
Finally, observe from \eqref{eq:lag-R} that we can compute,
for each $r=1,2,\ldots,R$, 
\[
\Lambda_r(x_0)=\frac{1}{(-1)^{R-r}F_{r-1}F_{R-r}}\cdot
               \frac{\Gamma(x_0)}{(x_0-r)}\pmod q\,.
\]
This initialization takes in total $O(R)$ operations and space.

By choosing a small enough $\epsilon>0$ to
subsume the polylogarithmic terms, the entire evaluation algorithm 
runs thus
in $O(N^{\omega+\epsilon/2})$ time and space. 
This completes the Camelot framework for computing
and verifying $X_{\binom{6}{2}}$ and thus proves 
Theorem~\ref{thm:k-clique-camelot}.

\section{A Camelot algorithm for the number of triangles}

\label{sect:triangle}

This section proves Theorem~\ref{thm:triangle-camelot}
and Theorem~\ref{thm:triangle} by showing that an algorithm design 
of Itai and Rodeh~\cite{Itai1978} for counting triangles in a graph 
can be transformed into a Camelot algorithm via the polynomial
extension of the split/sparse Yates's algorithm 
(\S\ref{sect:yates-poly}). 

\subsection{The Itai--Rodeh reduction to trace of a product matrix}

To count triangles in an $n$-vertex graph, it suffices to 
compute the matrix trace of the product $ABC$ of three $n\times n$ 
matrices $A,B,C$ with $\{0,1\}$ entries. 
Let us furthermore assume that each of these 
matrices has at most $m$ nonzero entries (that is, the graph has 
at most $m$ edges, and $A=B=C$ is the adjacency matrix of the graph).
In notation, we must compute $\sum_{i,j,k} a_{ij}b_{jk}c_{ki}$.

\subsection{The trace of a triple product of sparse matrices 
in parallel}

\label{sect:parallel-trace}

Let us start by proving Theorem~\ref{thm:triangle} and only
then develop the proof polynomial and the Camelot algorithm.

Let $\F$ be a field. For $r=1,2,\ldots,R$,
let $\alpha_{ij}(r),\beta_{jk}(r),\gamma_{ki}(r)$ be field
elements that satisfy the polynomial identity
\begin{equation}
\label{eq:trace-tri-consistency}
\sum_{i,j,k}a_{ij}b_{jk}c_{ki}\!
=
\!\sum_{r=1}^R
\biggl(
\sum_{i,j'}\alpha_{ij'}(r)a_{ij'}
\!\biggr)
\biggl(
\sum_{j,k'}\beta_{jk'}(r)b_{jk'}
\!\biggr)
\biggl(
\sum_{k,i'}\gamma_{ki'}(r)c_{ki'}
\!\biggr)\,.\!\!\!
\end{equation}
We can assume that $R=O(n^{\omega+\epsilon})$ for an arbitrary 
constant $\epsilon>0$, where $\omega$ is the limiting exponent 
for the tensor rank of the matrix multiplication tensor 
$\bra n,n,n\ket$ in $\F$~\cite{Burgisser1997,Landsberg2012}.%

Analogously to \S\ref{sect:small-clique-eval},
let us take a look at the detailed structure of the 
coefficients $\alpha_{ij}(r),\beta_{jk}(r),\gamma_{ki}(r)$.
We will focus on the coefficients
$\alpha_{ij}(r)$, the cases for $\beta_{jk}(r)$ and $\gamma_{ki}(r)$
are symmetric. 

By the properties of tensor rank and the fact that matrix 
multiplication tensors are closed under taking of Kronecker products, 
for any constant $\epsilon>0$ we can assume without loss 
of generality that $R=O(n^{\omega+\epsilon})$ and that there 
exist positive integer constants $R_0,n_0$ and a positive 
integer $t$ such that $R=R_0^t$ and $n=n_0^t$. In particular, 
we may identify the indices $d,e$ and the index $r$ with $t$-digit 
integers in base $n_0$ and $R_0$, respectively, so that if we write
$i_\ell,j_\ell$ and $r_\ell$ for the $\ell$th digit of 
$i,j$ and $r$, respectively,
the coefficient $\alpha_{ij}(r)$ has the Kronecker product form 
\begin{equation}
\label{eq:trace-alpha-kron}
\alpha_{ij}(r)=\prod_{\ell=1}^t\alpha^{(0)}_{i_\ell j_\ell}(r_\ell)\,,
\end{equation}
where $\alpha^{(0)}$ is a matrix of size $n_0^2\times R_0$ with 
integer entries. 
Thus, we may view \eqref{eq:trace-alpha-kron} as an integer 
matrix of size $n^2\times R$ which has been obtained as 
the $t$-fold Kronecker power of the constant-sized-and-entried 
matrix $\alpha^{(0)}$. 

It follows that we can use the split/sparse Yates algorithm 
(\S\ref{sect:yates-split-sparse}) to produce the values
\[
A_r=\sum_{i,j'}\alpha_{ij'}(r)a_{ij'}\,,\qquad r=1,2,\ldots,R\,,
\]
required in \eqref{eq:trace-tri-consistency} in parts of 
$O(m)$ values each. By the structure of the split/sparse algorithm, 
these $O(R/m)$ parts can be produced in parallel, 
independently of each other. By symmetry, the same applies
to the values 
\[
\begin{split}
B_r&=\sum_{j,k'}\beta_{jk'}(r)b_{jk'}\,,\qquad r=1,2,\ldots,R\,,\\
C_r&=\sum_{k,i'}\gamma_{ki'}(r)c_{ki'}\,,\qquad r=1,2,\ldots,R\,.
\end{split}
\]
Theorem~\ref{thm:triangle} now follows by working over the
rational numbers%
\footnote{Or sufficiently many distinct prime moduli to enable 
reconstruction using the Chinese Remainder Theorem.}{}
since we have 
$\sum_{i,j,k} a_{ij}b_{jk}c_{ki}=\sum_{r=1}^R A_rB_rC_r$
by \eqref{eq:trace-tri-consistency}.

\subsection{The proof polynomial and evaluation algorithm}

Let us now prove Theorem~\ref{thm:triangle-camelot} by extending
the algorithm in the previous section into a proof polynomial
and its evaluation algorithm. In essence, we replace the
split/sparse algorithm (\S\ref{sect:yates-split-sparse}) 
with its polynomial extension (\S\ref{sect:yates-poly}).
Let us assume that $\F=\Z_q$ is a finite field of prime order $q$.

Again it is convenient to restrict to consider the values
$A_r$, the values $B_r$ and $C_r$ are treated symmetrically. 

Let $m'$ be the minimum multiple of $R_0$ at least $m$. 
In particular, $m'$ divides $R$ and $m'=\Theta(m)$.
The polynomial extension of Yates's algorithm for the
coefficients $A_1,A_2,\ldots,A_R$ works with polynomials
$A_{r'}(z)$ of degree at most $R/m'$, where $r'=1,2,\ldots,m'$. 

Let $q$ be a prime at least $3R/m'+1$. 
We can find such a prime in time $\tilde O(1)$~\cite{Agrawal2004}.
For a given input $z_0\in\Z_q$, the polynomial extension 
algorithm produces in time $\tilde O(m'+R/m')$
the values $A_{1}(z_0),A_{2}(z_0),\ldots,A_{m'}(z_0)\pmod q$.
By the structure of the algorithm, 
for $z_0\in [R/m']$ and $r'=1,2,\ldots,m'$
these values constitute exactly the values $A_1,A_2,\ldots,A_R
\pmod q$. 

Pursuing the same approach for the values $B_r$ and $C_r$, 
we observe that the polynomials $A_{r'}(z),B_{r'}(z),C_{r'}(z)$ 
have the property that 
\begin{equation}
\label{eq:tri-product}
\sum_{z_0\in [R/m']}\sum_{r'=1}^{m'} A_{r'}(z_0)B_{r'}(z_0)C_{r'}(z_0)
=\sum_{r=1}^R A_rB_rC_r\,.
\end{equation}
Let us now define the proof polynomial
\[
P(z)=\sum_{r'=1}^{m'} A_{r'}(z)B_{r'}(z)C_{r'}(z)\,,
\]
and observe that it has degree at most $3R/m'$.

From \eqref{eq:tri-product} and \eqref{eq:trace-tri-consistency} 
we observe that
\[
\sum_{z_0\in [R/m']}P(z_0)=\sum_{i,j,k}a_{ij}b_{jk}c_{ki}\,.
\]
In particular, if we have $P(z)\pmod q$ in coefficient form, 
we can compute the trace of $ABC$ modulo $q$
in $\tilde O(R/m)$ operations.
We can recover $P(z)\pmod q$ in coefficient form from
at least $3R/m+1$ evaluations in $\tilde O(R/m)$ operations.
Each evaluation takes $\tilde O(m+R/m)$ operations. 
Since $A,B,C$ are $\{0,1\}$-valued matrices, the trace of
$ABC$ is $O(n^3)$. Up to polylogarithmic factors we may assume
that $q$ grows at least as fast as a root function of $n$.
Thus, we can use $O(1)$ distinct primes $q$ 
and the Chinese Remainder Theorem to recover 
the trace of $ABC$ from evaluations modulo $q$.
This gives proof size $\tilde O(R/m)$.
To complete the proof of Theorem~\ref{thm:triangle-camelot}, 
select a small enough $\epsilon>0$ to subsume the 
polylogarithmic factors.

\subsection{Meeting the Alon--Yuster--Zwick bound}

This section proves Theorem~\ref{thm:triangle-sparse}.
In essence, we observe that the Alon--Yuster--Zwick 
design~\cite{Alon1997} admits edge-linear parallel execution
if we implement the dense part of the algorithm using
the split/sparse variant of Yates's algorithm 
(\S\ref{sect:yates-split-sparse}).

Let $\epsilon>0$ be fixed and, with foresight, 
let $\Delta=m^{(\omega-1)/(\omega+1)}$. 
Without loss of generality we may assume that the input graph
is connected and hence the number of vertices $n$ satisfies
$m\geq n-1$. In time $\tilde O(m)$ we can compute the vertex
degrees and partition the vertices into {\em low} degree vertices
of degree at most $\Delta$ and {\em high} degree vertices of
degree above $\Delta$. It is immediate that there are at most
$2m/\Delta$ high-degree vertices. In time $\tilde O(m)$ we can 
compute the sparse representation of the graph induced by the
high-degree vertices. In particular, this induced subgraph
has at most $m/\Delta$ vertices and at most $m$ edges. 
Using the split/sparse algorithm on this subgraph 
as in \S\ref{sect:parallel-trace}
with $R=O\bigl((m/\Delta)^{\omega+\epsilon}\bigr)=O\bigl(m^{2(\omega+\epsilon)/(\omega+1)}\bigr)$, we can count the triangles 
consisting only of high-degree vertices in per-node time 
$\tilde O(m)$ using $O\bigl(m^{(\omega-1+\epsilon)/(\omega+1)}\bigr)$
parallel nodes. 

It remains to count the triangles with at least one low-degree
vertex. Label the edge ends incident to each low-degree 
vertex with unique numbers from 
$1,2,\ldots,\lfloor\Delta\rfloor$. 
Node $u=1,2,\ldots,\lfloor\Delta\rfloor$ now proceeds as 
follows. For each edge $e$ with end-vertices $x$ and $y$,
consider both ends $x$ and $y$ in turn. (We only describe
the process for $x$, the process for $y$ is symmetric.) 
If $x$ has low degree, let $z$ be the other end-vertex of the edge
whose $x$-end has been labeled with the label $u$ (if any). 
If $z$ and $y$ are adjacent (which we can check, e.g., by binary 
search to a sorted list of edges), we have found a triangle $x,y,z$. 
We let this triangle contribute to the count of triangles 
if both (a) $e$ is the (for example, lexicographic) minimum 
edge in the triangle $x,y,z$ with at least one low-degree end-vertex,
and (b) $x$ is the minimum low-degree vertex in $e$. 
It follows that we can count the triangles that have at least
one low-degree vertex in per-node time $\tilde O(m)$ 
using $O(\Delta)$ parallel nodes.

\section{A proof template for partitioning sum-products}

\label{sect:template}

This section supplies a template for proof polynomials that underlies 
our results for the chromatic and Tutte polynomials. 
We rely on Kronecker substitution to enable succinct univariate 
encoding of the partitioning property. 
(See Kedlaya and Umans~\cite{Kedlaya2011}
for a discussion of Kronecker substitution and its inverse.)

\subsection{The problem}

Let $U=\{1,2,\ldots,n\}$ be an $n$-element ground set (or {\em universe}),
and let us write $2^U$ for the set of all subsets of $U$.
Let us call a function $f:2^U\rightarrow\Z$ a {\em set function}. 

Suppose the common input consists of at least $n$ bits, known to all 
the nodes. Suppose the input defines a set function $f$ such that for any
given $X\subseteq U$ any node can compute the value $f(X)$ in time $O^*(1)$.
{\em Furthermore, we assume that based on the input each node can 
in $O^*(1)$ time compute an upper bound $\phi$ such that $|f(X)|\leq\phi$
for all $X\subseteq U$.}

For a positive integer $t$ with $t=O^*(1)$, the nodes would
like to compute the value of the {\em $t$-part partitioning sum product}
\begin{equation}
\label{eq:p-part}
\sum_{(X_1,X_2,\ldots,X_t)} f(X_1)f(X_2)\cdots f(X_t)
\,,
\end{equation}
where the sum is over all $t$-tuples 
$(X_1,X_2,\ldots,X_t)\in 2^U\times 2^U\times\cdots\times 2^U$ such that 
\begin{equation}
\label{eq:partition}
X_1\cup X_2\cup \cdots X_t=U
\text{ and $X_i\cap X_j=\emptyset$ holds for all $1\leq i<j\leq t$}.
\end{equation}
(That is, the $t$-tuple $(X_1,X_2,\ldots,X_t)$ partitions the 
universe $U$ into $t$ pairwise disjoint {\em parts} $X_1,X_2,\ldots,X_t$ 
so that the ordering of the parts is relevant and zero or more empty parts 
are permitted.)

\medskip
\noindent
{\em Remark.}
On its own, a node can compute \eqref{eq:p-part} in time 
$O^*(2^n)$~\cite{BHK2009}.

\subsection{The proof polynomial}

This section defines the proof polynomial $P(x)$ 
for partitioning sum-products. 

We start by partitioning the universe $U$ into two disjoint parts, 
$U=E\cup B$, where $E$ is the {\em explicit} set (whose subsets each 
node will track explicitly) and $B$ is, by slight abuse of terminology, 
a set of {\em bits} (where the node will rely on advice from the other nodes).

Let us assume that the set $B$ of bits is 
$B=\{0,1,\ldots,2^{|B|-1}\}$. 
With this assumption, select exactly $|B|$ bits, possibly with 
repetition. Observe that if your bits sum to $2^{|B|}-1$, 
then you have selected each bit exactly once (without repetition). 

Let us also recall that there are exactly $\binom{m+k-1}{k-1}$ ways to place
$m$ identical balls into $k$ distinct bins. Put otherwise, we can form
exactly $\binom{m+k-1}{k-1}$ distinct multisets of size $m$ over
a ground set of $k$ elements. Let us write $\multiset{B}{k}$ for the 
set of multisets of size $k$ over $B$. In particular, 
$|\multiset{B}{|B|}|=\binom{2|B|-1}{|B|-1}$ is precisely the number
of ways we can select $|B|$ bits out of $B$, possibly with repetition.
Let us write $\sum M$ for the sum of elements in $M\in\multiset{B}{|B|}$.
With this notation, our previous observation is equivalent the
statement that for all $M\in\multiset{B}{|B|}$ we have 
$\sum M=2^{|B|}-1$ if and only if $M=B$.

It will be convenient to present the proof polynomial first
over the integers, and only then choose appropriate primes $q$.
The proof polynomial is
\begin{equation}
\label{eq:part-proof}
P(x)=p_0+p_1 x+p_2x^2+\ldots+p_d x^d\, 
\end{equation}
whose coefficients are defined for all $s=0,1,\ldots,d$ with 
$d=2^{|B|-1}|B|$ by
\begin{equation}
\label{eq:proof}
p_s=
\sum_{(X_1,X_2,\ldots,X_t)}f(X_1)f(X_2)\cdots f(X_t)\,,
\end{equation}
where the sum is over all $t$-tuples 
$(X_1,X_2,\ldots,X_t)\in 2^U\times 2^U\times\cdots\times 2^U$ that 
satisfy the multiset equality
\begin{equation}
\label{eq:constraint}
X_1+X_2+\ldots+X_t= E+M
\end{equation}
for a multiset $M\in\multiset{B}{|B|}$ with $\sum M=s$.
(Here we use additive notation for multisets to stress that the element 
multiplicities must agree for each element in the ground set $U$.)

\medskip
\noindent
{\em Remark.}
We observe that $p_{2^{|B|-1}}$ agrees with \eqref{eq:p-part}. 
Indeed, \eqref{eq:partition} holds if and only if
\eqref{eq:constraint} holds with $B=M$. 
Furthermore, $B=M$ if and only if $\sum M = 2^{|B|-1}$.
Thus, each node needs to get confidence
that the coefficients $p_0,p_1,\ldots,p_d$ have been 
correctly computed, {\em over the integers}. 

\medskip
\noindent
{\em Remark 2.}
Since $P(x)$ has degree at most $2^{|B|-1}|B|$, it suffices
to use primes $q$ with $q=O^*(2^{|B|})$ to enable evaluation
and reconstruction of the proof modulo $q$. 
Such primes can be found in time $O^*(1)$~\cite{Agrawal2004}.

\medskip
\noindent
{\em Remark 3.}
Each node can use the upper bound $\phi$ for the absolute values of 
the function $f$, to compute in time $O^*(1)$ the upper bound 
$2^{nt+1}\phi^t$ for the absolute values in \eqref{eq:proof}.
Assuming that $\log q=\Omega(n)$ and $\log q=\Omega(\phi)$, 
which will be the case in our instantiations of the template, 
it suffices to work with $O^*(1)$ distinct primes $q$ to reconstruct 
the {\em integer} coefficients $p_0,p_1,\ldots,p_d$. Indeed, recall 
that $t=O^*(1)$ and that $\log\phi=O^*(1)$ since $\phi$ is 
computable from the common input in time $O^*(1)$.

\subsection{A template for the evaluation algorithm}

Let $x_0\in\{0,1,\ldots,q-1\}$ be given. 
We now describe how a node computes $P(x_0)\pmod q$, but will leave 
one implementation detail unspecified, namely how the function $g$ is 
computed within the desired computational budget, with the understanding 
that such algorithms will be provided in subsequent sections when the template 
is instantiated.

Let $w_E,w_B$ be formal indeterminates. (That is, each node will be computing 
with polynomials in $w_E,w_B$ with integer coefficients normalized to
$\{0,1,\ldots,q-1\}$ modulo $q$.)
Each node computes the function% 
\footnote{By ``computing the function $g$'' we mean that the node 
prepares a complete table of values of the function for every 
possible input $Y\subseteq E$.}
$g:2^E\rightarrow\Z_q[w_E,w_B]$ 
defined for all $Y\subseteq E$ by
\begin{equation}
\label{eq:node-function}
g(Y)=\sum_{\substack{X\subseteq U\\X\cap E\subseteq Y}}f(X)w_E^{|X\cap E|}w_B^{|X\cap B|}\prod_{b\in X\cap B}x_0^b\pmod q\,.
\end{equation}
(Recall that $B=\{0,1,\ldots,2^{|B|-1}\}$ so the expression $x_0^b$ is 
well-defined.)

\medskip
\noindent
{\em Remark.} The time budget for computing the function
$g$ is $O^*(2^{|E|})$. This will be justified on a case-by-case basis
in subsequent sections.

\medskip
\noindent
Assuming the node has computed the function $g$, it next 
computes
\begin{equation}
\label{eq:node-sieve}
a(w_E,w_B) = \sum_{j_E,j_B} a_{j_E,j_B} w_E^{j_E}w_B^{j_B}
     = \sum_{Y\subseteq E}(-1)^{|E\setminus Y|}g(Y)^t\pmod q\,.
\end{equation}
The node is interested on $a_{|E|,|B|}$, the integer coefficient 
of the monomial $w^{|E|}_Ew_B^{|B|}$ in $a(w)$. 
Using \eqref{eq:node-sieve} and the computed $g$ directly, 
the node can compute $a_{|E|,|B|}$ in time $O^*(2^{|E|})$.

From \eqref{eq:node-function}, \eqref{eq:node-sieve}, and 
standard inclusion--exclusion arguments \cite{BHK2009} it follows that
\begin{equation}
\label{eq:node-coefficient}
a_{|E|,|B|}=
\sum_{(X_1,X_2,\ldots,X_t)}
f(X_1)f(X_2)\cdots f(X_t)
\prod_{j=1}^t\prod_{b\in X_j\cap B} x_0^b\pmod q\,,
\end{equation}
where the sum is over all $t$-tuples 
$(X_1,X_2,\ldots,X_t)\in 2^U\times 2^U\times\cdots\times 2^U$ 
for which there exists a multiset $M\in\multiset{B}{|B|}$
with  
\begin{equation}
\label{eq:node-constraint}
X_1+X_2+\ldots+X_t=E+M\,.
\end{equation}
Furthermore, every multiset $M\in\multiset{B}{|B|}$ satisfies 
$\sum M\in\{0,1,\ldots,2^{|B|-1}|B|\}$.

Comparing \eqref{eq:proof} and \eqref{eq:constraint} with 
\eqref{eq:node-coefficient} and \eqref{eq:node-constraint}, we observe that 
\[
P(x_0)=a_{|E|,|B|}\pmod q\,.
\]
That is, we have evaluated the polynomial $P(x)$ at the chosen 
point $x_0$ modulo $q$.

\subsection{Running time for evaluation}

Assuming the node keeps to its budget for computing the function
$g$, we observe that $P(x_0) \pmod q$ for a given $x_0\in\{0,1,\ldots,q-1\}$
can be computed in time $O^*(2^{|B|}+2^{|E|})$.
This is minimized when $|B|=|E|$. Since $n=|U|=|E|+|B|$,
we have $|E|=|B|=n/2$.%

\section{Counting exact set covers}

\label{sect:set-covers}

As a warmup to illustrate the template in \S\ref{sect:template}, 
let us instantiate it for the problem of counting exact set covers.%

\subsection{Counting exact set covers via partitioning sum-products}

Suppose the input a set $\mathcal{F}$ of $O^*(2^{n/2})$ subsets of
an $n$-element universe $U$. Our task is to compute the number 
of distinct set partitions of $U$ that we can form by using exactly $t$ 
sets from $\mathcal{F}$, for a given $1\leq t\leq n$.%
\footnote{To avoid degenerate cases, let us assume that $\mathcal{F}$
does not contain the empty set.}

To instantiate the template, let us take the set function 
$f:2^{U}\rightarrow\Z$ to be the indicator function for 
the sets in $\mathcal{F}$.
That is, for all $X\subseteq U$ we define
\begin{equation}
\label{eq:f-indicator}
f(X)=\begin{cases}
1 &\text{if $X\in\mathcal{F}$};\\
0 & \text{otherwise}.
\end{cases}
\end{equation}
In particular, we can take $\phi=1$.
Then, the value of the partitioning sum-product \eqref{eq:p-part} 
is exactly $t!$ times our desired solution. 

\subsection{Computing the node function}

It remains to complete the template for evaluating the 
function \eqref{eq:node-function} with $f$ defined 
by \eqref{eq:f-indicator} in time $O^*(2^{n/2})$. 
In particular, we instantiate the template with $|E|=|B|=n/2$.

\medskip
\noindent
{\em Remark.}
Computing \eqref{eq:node-function}
within the time budget requires a dedicated algorithm since 
$\mathcal{F}$ may have $O^*(2^{n/2})$ sets, and hence the sums 
in \eqref{eq:node-function} (across all choices $Y\subseteq E$) 
may have up to $2^{n}$ terms in total, which is well above our 
time budget.

The node proceeds as follows. Let $q$ and $x_0\in\{0,1,\ldots,q-1\}$
be as in the template. 
Initialize an array $g_0$ whose entries are indexed by subsets in $2^E$.
Each entry $g_0(Z)$ for $Z\subseteq E$ will be a polynomial in 
the indeterminates $w_E,w_B$ and with coefficients in $\Z_q$. 
We assume that initially $g_0(Z)=0$ for all $Z\subseteq E$. 
Next, we iterate over the $O^*(2^{n/2})$ sets in $\mathcal{F}$.
For each $X\in\mathcal{F}$, set $g(X\cap E)\leftarrow g(X\cap E)
w_E^{|X\cap E|}w_B^{|X\cap B|}x_0^{\sum X\cap B}\pmod q$.
This iteration can be implemented to run in time $O^(2^{n/2})$.
After the iteration is complete, use Yates's algorithm \cite{Yates1937} 
on $g_0$ to obtain in time $O^*(2^{n/2})$ the function $g$ with 
\[
g(Y)=\sum_{Z\subseteq Y}g_0(Z)=
\sum_{\substack{X\in\mathcal{F}\\X\cap E\subseteq Y}}
w_E^{|X\cap E|}w_B^{|X\cap B|}x_0^{\sum X\cap B}\pmod q\,.
\]
Thus, the function \eqref{eq:node-function} 
can be computed in time $O^*(2^{n/2})$.

\section{The chromatic polynomial}

\label{sect:chromatic}

Let us now proceed with a somewhat more intricate application. 
We instantiate the template in \S\ref{sect:template} 
for the chromatic polynomial. 

\subsection{The chromatic polynomial via partitioning sum-products}

Let $G$ be an undirected graph with $n$ vertices. 
The chromatic polynomial $\chi_G(t)$ is a polynomial in $t$ 
of degree at most $n$. Thus the values of $\chi_G(t)$ at
any $n+1$ points suffice to reconstruct $G$ by interpolation.

The value $\chi_G(t)$ for a positive integer $t=1,2,\ldots,n+1$ 
counts the number of mappings $c:V(G)\rightarrow \{1,2,\ldots,t\}$ such
that $c(u)\neq c(v)$ holds for all edges $\{u,v\}\in E(G)$.
Such mappings are in a bijective correspondence with the
$t$-tuples $(X_1,X_2,\ldots,X_t)$ that partition $U=V(G)$
into $t$ pairwise disjoint parts $X_1,X_2,\ldots,X_t$ such 
that each $X_i\subseteq U$ is an independent set of $G$.
Indeed, take $X_k=c^{-1}(k)$ for $k=1,2,\ldots,t$ to see
the correspondence between preimages of $c$ and the parts. 

Thus, $\chi_G(t)$ equals the partitioning sum-product \eqref{eq:p-part}
if we choose the set function $f:2^{U}\rightarrow\Z$ 
to be the indicator function for independent sets in $G$.
That is, for all $X\subseteq U$ we define
\begin{equation}
\label{eq:f-independent}
f(X)=\begin{cases}
1 &\text{if $X$ is independent in $G$};\\
0 & \text{otherwise}.
\end{cases}
\end{equation}
In particular, we can take $\phi=1$ and instantiate the template
for $t=1,2,\ldots,n+1$ to recover $\chi_G(t)$ 
from $\chi_G(1),\chi_G(2),\ldots,\chi_G(n+1)$ by polynomial
interpolation.

\subsection{Computing the node function}

Suppose that $t=1,2,\ldots,n+1$ is fixed.
It remains to complete the template for evaluating the 
function \eqref{eq:node-function} with $f$ defined 
by \eqref{eq:f-independent} in time $O^*(2^{n/2})$. 
In particular, we instantiate the template with $|E|=|B|=n/2$.

\medskip
\noindent
{\em Remark.}
Computing \eqref{eq:node-function}
within the time budget requires a dedicated algorithm since $G$ may have
up to $2^n$ independent sets, and hence the sum \eqref{eq:node-function}
may have up to $2^n$ terms (for $Y=E$), which is well above our
time budget.

\medskip

The intuition that we pursue in what follows is that an arbitrary 
independent set $I\subseteq E\cup B$ in $G$ remains independent in $G$ 
when restricted to the induced subgraphs $G[E]$ and $G[B]$. 
Furthermore, $I\cap B$ is disjoint from the neighborhood of $I\cap E$ 
in $B$ {\em but otherwise has no interactions with $I\cap E$}. 
These observations enable us to compute \eqref{eq:node-function} 
in parts across the cut $(E,B)$ with the help of Yates's 
algorithm~\cite{Yates1937}. 

The node proceeds as follows. As per the template, the node first selects
a prime $q$ and a uniform random integer $x\in\{0,1,\ldots,q-1\}$. 
Then, the node computes the function 
$\hat f_B:2^B\rightarrow\Z_q[w_E,w_B]$
defined for all $X\subseteq B$ by
\[
f_B(X)=
\begin{cases}
\prod_{b\in X}w_Bx_0^b & \text{if $X$ is independent in $G$};\\
0 & \text{otherwise}
\end{cases}
\pmod q\,.
\]
This takes time $O^*(2^{|B|})$. 
Next, using Yates's algorithm \cite{Yates1937}, 
the node computes in time $O^*(2^{|B|})$ 
the function $g_B:2^B\rightarrow\Z_q[w_E,w_B]$
defined for all $Y\subseteq B$ by
\[
g_B(Y)=\sum_{X\subseteq Y} f_B(X)\pmod q\,.
\]

For a subset $X\subseteq E$, let us write $\Gamma_{G,B}(X)$ for
the set of all vertices $v\in B$ that have at least one neighbor
in the set $X$ in $G$. 
Next, using the function $g_B$ that the node has computed, 
it computes the function 
$\hat f_E:2^E\rightarrow\Z_q[w_E,w_B]$
defined for all $X\subseteq E$ by
\begin{equation}
\label{eq:hat-f-e}
\hat f_E(X)=\begin{cases}
w_E^{|X|}g_B\bigl(B\setminus\Gamma_{G,B}(X)\bigr) & 
\text{if $X$ is independent in $G$};\\
0 & \text{otherwise}
\end{cases}
\pmod q\,.
\end{equation}
\noindent
{\em Remark.}
The term $g_B\bigl(B\setminus\Gamma_{G,B}(X)\bigr)$ in \eqref{eq:hat-f-e} 
extends an independent set $X\subseteq E$ with contributions from 
independent sets in $B$ in all possible ways that are compatible with $X$.
This aggregation of contributions across the cut $(E,B)$ is the key 
to staying within the time budget. 

\medskip
Finally, the node computes the function $g$ in time $O^*(2^{|E|})$
using the function $\hat f_E$ and Yates's algorithm. Indeed, 
for all $Y\subseteq E$ by comparing computations
\eqref{eq:node-function}, and \eqref{eq:f-independent}, 
we observe that
\[
g(Y)=\sum_{X\subseteq Y}\hat f_E(X)\pmod q\,.
\]

\section{The Tutte polynomial}

\label{sect:tutte}

We instantiate the template in \S\ref{sect:template}
for the Tutte polynomial. 

\subsection{The Tutte polynomial via partitioning sum-products}

Let $G$ be an undirected graph with $n$ vertices and $O^*(1)$ edges,
possibly including loops and parallel edges. 
First, let us recall 
that task of computing the Tutte polynomial $T_G(x,y)$ 
can be reduced to the task of evaluating the partition function of 
the $t$-state Potts model with integer edge weights. 

Towards this end, associate with $G$ the multivariate polynomial
\[
Z_G(t,r)=\sum_{F\subseteq E} t^{c(F)}\prod_{e\in F} r_e\,,
\]
where $t$ is an indeterminate, $r$ is a vector of indeterminates,
with one indeterminate $r_e$ for each edge $e\in E=E(G)$,
and $c(F)$ denotes the number of connected components in the
subgraph of $G$ with vertex set $V$ and edge set $F$. 

The Tutte polynomial $T_G(x,y)$ can be recovered from $Z_G(t,r)$ by
observing \cite{Sokal2005} that 
\begin{equation}
\label{eq:tutte-potts}
T_G(x,y)=(x-1)^{-c(E)}(y-1)^{-|V|}Z_G(t,r)\,,
\end{equation}
where $t=(x-1)(y-1)$ and $r_e=y-1$ for all $e\in E$. 

For integer values $t=1,2,\ldots$, Fortuin and Kasteleyn~\cite{Fortuin1972} 
observed that
\[
Z_G(t,r)=\sum_{\sigma:V\rightarrow\{1,2,\ldots,t\}}\prod_{e\in E}\bigl(1+r_e[\sigma(e_1)=\sigma(e_2)]\bigr)\,.
\]
(Here we write $e_1$ and $e_2$ for the end-vertices of $e\in E$.)
Next, assume that $r_e=r$ for all $e\in E$. 
Define the function $f:2^{U}\rightarrow\Z$ for all 
$X\subseteq U=V(G)$ by
\begin{equation}
\label{eq:potts-inner}
f(X)=(1+r)^{|E(G[X])|}\,.
\end{equation}
For integer values $r=1,2,\ldots$ it holds that
\begin{equation}
\label{eq:potts-sum}
Z_G(t,r)=\sum_{(X_1,X_2,\ldots,X_t)}f(X_1)f(X_2)\cdots f(X_t)\,.
\end{equation}
where the sum is over all $t$-tuples 
$(X_1,X_2,\ldots,X_t)\in 2^U\times 2^U\times\cdots\times 2^U$ such that 
\[
X_1\cup X_2\cup \cdots X_t=U
\text{ and $X_i\cap X_j=\emptyset$ holds for all $1\leq i<j\leq t$}.
\]

Thus, to compute the Tutte polynomial of a graph $G$ it suffices
to compute the partitioning sum-product \eqref{eq:potts-sum} with the
inner function \eqref{eq:potts-inner} for $O^*(1)$ integer points 
$(t,r)$ with $t=O^*(1)$ and $r=O^*(1)$ to enable interpolation of 
the Tutte polynomial via \eqref{eq:tutte-potts}. 
In particular, we can take $\phi=O^*(1)$.

\subsection{Computing the node function}

Here we assume that $|E|=2|B|$.
Let $q$ be the chosen prime and $x_0\in\{0,1,\ldots,q-1\}$.
Let $t=O^*(1)$ and $r=O^*(1)$.
Let us clean up the node function $g$ somewhat.
To enable an application of Yates's algorithm to get the
final node function $g$, each node needs to compute 
the function $g_0:2^E\rightarrow\Z_q[w_E,w_B]$ 
defined for all $Y\subseteq E$ by
\[
g_0(Y)=
w_E^{|Y|}
\sum_{X\subseteq B}
w_B^{|X|}
x_0^{\sum X}
f(X\cup Y)\,,
\pmod q
\]
where the inner function $f$ is defined by 
\[
f(X\cup Y)=\prod_{e\in E(G[X\cup Y])}(1+r)=(1+r)^{|E(G[X\cup Y])|}\,.
\]

The inner function $f$ in particular involves interactions 
between $X$ and $Y$ across the cut $(E,B)$, which appear less 
easy to control/aggregate than similar interactions in the case 
of independent sets. Each node will rely on the following tripartite 
strategy to compute $g_0$.
Split $E$ into two disjoint subsets $E_1,E_2$ with $|E_1|=|E_2|$.
Split $Y\subseteq E$ accordingly into $Y_1=Y\cap E_1$ and
$Y_2=Y\cap E_2$. Since the interactions in $f$ are only between
at most two out of the three parts $E_1,E_2,B$, we have the product
decomposition
\[
f(X\cup Y_1\cup Y_2)=
f_{B,E_1}(X\cup Y_1)
f_{B,E_2}(X\cup Y_2)
f_{E_1,E_2}(Y_1\cup Y_2)\,,
\]
where
\[
\begin{split}
f_{B,E_1}(X\cup Y_1)&=(1+r)^{|E(G[X,Y_1])|+|E(G[X])|}\,,\\
f_{B,E_2}(X\cup Y_2)&=(1+r)^{|E(G[X,Y_2])|+|E(G[Y_2])|}\,,\\
f_{E_1,E_2}(Y_1\cup Y_2)&=(1+r)^{|E(G[Y_1,Y_2])|+|E(G[Y_1])|}\,.
\end{split}
\]
Let us now augment this decomposition by subsuming elements
from the node function $g_0$ to the terms of the decomposition.
Let 
\[
w_E^{|Y_1|+|Y_2|}w_{B}^{|X|}x_0^{\sum X}f(X\cup Y_1\cup Y_2)=
\hat f_{B,E_1}(X\cup Y_1)
\hat f_{B,E_2}(X\cup Y_2)
f_{E_1,E_2}(Y_1\cup Y_2)\,,
\]
where
\[
\begin{split}
\hat f_{B,E_1}(X\cup Y_1)&=(1+r)^{|E(G[X,Y_1])|+|E(G[X])|}w_E^{|Y_1|}w_{B}^{|X|}x_0^{\sum X}\,,\\
\hat f_{B,E_2}(X\cup Y_2)&=(1+r)^{|E(G[X,Y_2])|+|E(G[Y_2])|}w_E^{|Y_2|}\,,\\
f_{E_1,E_2}(Y_1\cup Y_2)&=(1+r)^{|E(G[Y_1,Y_2])|+|E(G[Y_1])|}\,.
\end{split}
\]
Assuming that $|E_1|=|E_2|=|B|$, each node can in time
$O^*(2^{|E|})$ compute each of the functions
$\hat f_{B,E_1}$, $\hat f_{B,E_2}$, and $f_{E_1,E_2}$.

For all $Y_1\subseteq E_1$ and $Y_2\subseteq E_2$ we have
\begin{equation}
\label{eq:node-g0-mm}
\begin{split}
g_0(Y_1\cup Y_2)
&=
f_{E_1,E_2}(Y_1\cup Y_2)
\sum_{X\subseteq B}
\hat f_{B,E_1}(X\cup Y_1)
\hat f_{B,E_2}(X\cup Y_2)\\
&=
f_{E_1,E_2}(Y_1\cup Y_2)
t_{E_1,E_2}(Y_1\cup Y_2)\,,
\end{split}
\end{equation}
where the function $t_{E_1,E_2}:2^{E_1\cup E_2}\rightarrow\Z_q[w_E,w_B]$
is defined for all $Y_1\subseteq E_1$ and $Y_2\subseteq E_2$ by
\begin{equation}
\label{eq:node-t}
t_{E_1,E_2}(Y_1\cup Y_2)=
\sum_{X\subseteq B}
\hat f_{B,E_1}(X\cup Y_1)
\hat f_{B,E_2}(X\cup Y_2)\,.
\end{equation}
In particular, $t_{B,E_1}$ is just a matrix product
of the matrices $\hat f_{B,E_1}$ and $\hat f_{B,E_2}$. 

Using \eqref{eq:node-g0-mm} and \eqref{eq:node-t}, 
for any constant $\epsilon>0$ each node
can compute the function $g_0$ in time 
$O^*(2^{|E_1|(\omega+\epsilon)/3})=O^*(2^{|E|(\omega+\epsilon)/2})=
O^*(2^{(\omega+\epsilon)n/3})$.
(Indeed, recall that we assume $|E_1|=|E_2|=|B|$.)
Within the same time budget, a node can recover $g$ by taking
the zeta tranform of $g_0$.

\section*{Acknowledgments}

We thank Ryan Williams for useful discussions and giving us
early access to his manuscript describing his batch evaluation
framework~\cite{Williams2016} of which the present ``Camelot''
framework forms a special case. We also thank Stefan Schneider and 
Russell Impagliazzo for an inspiring exposition of the 
nondeterministic exponential time hypothesis and the 
Carmosino {\em et al.} \cite{Carmosino2016} results at the 
Simons Institute in fall 2015.

\clearpage

\begin{center}
\huge
\textsc{Appendix}
\end{center}

\appendix

\section{An inventory of earlier polynomials}

\label{sect:pre-existing}

These polynomials are essentially due to Williams~\cite{Williams2016}, 
except for the set cover polynomial $F_t$, which is implicit 
in \cite{BHK2009}, and for the polynomial for 
Convolution3SUM~\cite{Patrascu2010}, 
which is a minor technical extension of the polynomials due 
to Williams.
In particular, we make no claims of originality for the polynomials
in this section but rather present the polynomials to 
(a) demonstrate the versatility of the Camelot framework, and 
(b) provide an accessible path towards the main results of the paper.

\subsection{A polynomial for counting Boolean orthogonal vectors}

\label{sect:ortho}

This section proves Theorem~\ref{thm:ortho-hamming-conv3sum}(1).
Let us recall the Boolean orthogonal vectors problem. We are 
given two matrices, $A=(a_{ij})$ and $B=(b_{ij})$, both of 
size $n\times t$ and with $\{0,1\}$ entries.
{\em For each of the rows in $A$, we want to count
the number of rows in $B$ orthogonal to it.}
In notation, we want to find, for each $i=1,2,\ldots,n$,
the count
\[
c_i=\sum_{k=1}^{n} \biggl[\sum_{j=1}^t a_{ij}b_{kj}=0\biggr]\,,
\qquad 0\leq c_i\leq n\,.
\]

Let us now define the proof polynomial $P(x)$. We will work
modulo a prime $q=\tilde O(nt)$. Using fast interpolation
(see \S\ref{sect:fast-poly}), in time $\tilde O(nt)$ we can 
find $t$ univariate polynomials $A_1(x),A_2(x),\ldots,A_t(x)$ of 
degree at most $n$ such that $A_j(i)=a_{ij}\pmod q$ holds for 
all $i=1,2,\ldots,n$ and $j=1,2,\ldots,t$. Let
\[
A(x)=(A_1(x),A_2(x),\ldots,A_t(x))\,.
\]
In time $\tilde O(nt)$ we can evaluate a $t$-variate
polynomial $B(z_1,z_2,\ldots,z_t)$ such that 
\[
B(z_1,z_2,\ldots,z_t)=
\sum_{i=1}^{n}\biggl[\sum_{j=1}^t b_{ij}z_j=0\biggr]\pmod q
\]
holds for all $(z_1,z_2,\ldots,z_t)\in\{0,1\}^t$.
Indeed, we can evaluate via the formula
\begin{equation}
\label{eq:b-orthogonal}
B(z_1,z_2,\ldots,z_t)=\sum_{i=1}^n\prod_{j=1}^t(1-b_{ij}z_j)\pmod q\,.
\end{equation}
Let us now take
\[
P(x)=B(A(x))\pmod q\,.
\]
To evaluate $P(x)$ at a point $x_0\in\Z_q$, 
a node first computes the vector $A(x_0)\in\Z_q^t$ and 
then uses the evaluation formula \eqref{eq:b-orthogonal} 
to obtain $P(x_0)$ in time $\tilde O(nt)$.
We observe that $P(x)$ has degree $d\leq nt$. 
Furthermore, for all $i=1,2,\ldots,n$ we have
\[
P(i)=B(a_{i1},a_{i2},\ldots,a_{it})=c_i\pmod q\,.
\] 
Since $0\leq c_i\leq n$, from $P(i)\pmod q$ we can recover the integer $c_i$.
From a correct proof $p_0,p_1,\ldots,p_d\in\Z_q$ 
we can by interpolation in time $\tilde O(q)$ thus recover the 
integer counts $c_1,c_2,\ldots,c_n$. 

\subsection{A polynomial for \#CNFSAT}

This section proves Theorem~\ref{thm:cnf-per-hp}(1).
Let the common input be a
CNF formula with $v$ variables $x_1,x_2,\ldots,x_v$ 
and $m$ clauses $C_1,C_2,\ldots,C_m$. 
For convenience, let us assume that $v$ is even.

In time $O^*(2^{v/2})$ each node can prepare two 
matrices $A=(a_{ij})$ and $B=(b_{ij})$, both of 
size $2^{v/2}\times m$, with the following structure.

For $i=1,2,\ldots,2^{v/2}$ and $j=1,2,\ldots,m$,
the entry $a_{ij}=1$ if and only if 
the $i$th Boolean assignment to variables $x_1,x_2,\ldots,x_{v/2}$
satisfies {\em no} literal in $C_j$; otherwise $a_{ij}=0$.
The entry $b_{ij}=1$ if and only if 
the $i$th Boolean assignment to variables $x_{v/2+1},x_{v/2+2},\ldots,x_{v}$
satisfies {\em no} literal in $C_j$; otherwise $b_{ij}=0$.

Observe now that for a Boolean assignment $(i_1,i_2)\in [2^{v/2}]^2$
to the variables $x_1,x_2,\ldots,x_v$ we have
$\sum_{j=1}^m a_{i_1j}b_{i_2j}=0$ if and only if 
$(i_1,i_2)$ satisfies {\em} all the clauses. 

We can thus reduce the task of counting all the satisfying assignments
to the task of counting orthogonal pairs of vectors (see~\S\ref{sect:ortho}), 
with $n=2^{v/2}$ and $t=m$. Since $m=O^*(1)$, we obtain a
Camelot algorithm for \#CNFSAT that prepares a proof of 
size $O^*(2^{v/2})$ in time $O^*(2^{v/2})$.

\subsection{A polynomial for the Hamming distance distribution}

Now that we have had some initial exposure to the framework, let us
upgrade the Boolean orthogonal vectors setup to produce the entire 
Hamming distance distribution over the $n^2$ pairs of vectors,
localized to each of the $n$ vectors in one set of vectors. 
The technical gist is that we can control the roots of a
(Lagrange) interpolating polynomial by supplying them via
separate indeterminates $w_1,w_2,\ldots,w_t$ 
to extract desired features from the input. 
This section proves Theorem~\ref{thm:ortho-hamming-conv3sum}(2).

We are given as input given two matrices, $A=(a_{ij})$ and $B=(b_{ij})$, 
both of size $n\times t$ and with $\{0,1\}$ entries.
{\em For each of the rows in $A$, we want to count, for
each possible Hamming distance $h$, the number of rows in $B$ 
with distance $h$ to the row.}
In notation, we want to find, {\em for each} $i=1,2,\ldots,n$
and $h=0,1,\ldots,t$, the count
\[
c_{ih}=\sum_{k=1}^{n} \biggl[\sum_{j=1}^t(1-a_{ij})b_{kj}+a_{ij}(1-b_{kj})=h\biggr]\,,
\qquad 0\leq c_{ih}\leq n\,.
\]

Let us now define the proof polynomial $P(x)$.
We will work modulo a prime $q=\tilde O(nt^2)$.
In time $\tilde O(nt^2)$ we can find $t$ univariate polynomials 
$A_1(x),A_2(x),\ldots,A_t(x)$ of degree at most $n(t+1)$ such that
$A_j(i(t+1)+h)=a_{ij}\pmod q$ holds for all $i=1,2,\ldots,n$, $h=0,1,\ldots,t$ 
and $j=1,2,\ldots,t$. 

Also in time $\tilde O(nt^2)$ we can find $t$ univariate polynomials 
$H_1(x),H_2(x),\ldots,H_t(x)$ of degree at most $n(t+1)$ 
such that for all $i=1,2,\ldots,n$ and $h=0,1,\ldots,t$ it holds that
\[
\bigl\{H_1\bigl(i(t+1)+h\bigr),
       H_2\bigl(i(t+1)+h\bigr),
       \ldots,
       H_t\bigl(i(t+1)+h\bigr)\bigr\}=
\{0,1,\ldots,t\}\setminus\{h\}\,.
\]
\noindent
{\em Remark.} Above it does not matter which polynomial $H_j$ attains 
which of the $t$ values in $\{0,1,\ldots,t\}\setminus\{h\}$, but this 
precise set of $t$ values must be correctly attained for each $i$. 
For example, we can force the polynomials to take values so that 
for $j=1,2,\ldots,t$ each polynomial $H_j$ takes the lowest value 
$\{0,1,\ldots,t\}\setminus\{h\}$ not already attained.
\medskip

Let
\[
I(x)=\bigl(A_1(x),A_2(x),\ldots,A_t(x),H_1(x),H_2(x),\ldots,H_t(x)\bigr)\,.
\]
In time $\tilde O(nt)$ we can evaluate a $(t+t)$-variate
polynomial 
\[
B(z_1,z_2,\ldots,z_t,w_1,w_2,\ldots,w_t)
\]
of degree at most $t$ such that 
for all $(z_1,z_2,\ldots,z_t)\in\{0,1\}^t$ and $h=0,1,\ldots,t$ 
we have%
\footnote{Here we use the somewhat sloppy notation
$\{0,1,\ldots,t\}\setminus\{h\}$ to indicate that we assign
these $t$ values in arbitrary order to the $t$ variables
$w_1,w_2,\ldots,w_t$ of $B$.}{}
\[
\begin{split}
B(z_1,&z_2,\ldots,z_t,\{0,1,\ldots,t\}\setminus\{h\})
=\\
&\bigl(\Pi_{\substack{\ell=0\\\ell\neq h}}^t(h-\ell)\bigr)
\sum_{i=1}^{n}\biggl[\sum_{j=1}^t (1-z_j)b_{ij}+z_j(1-b_{ij})=h\biggr]\pmod q\,.
\end{split}
\]
Indeed, we can evaluate via the formula
\begin{equation}
\label{eq:b-hamming}
\begin{split}
B(z_1,&z_2,\ldots,z_t,w_1,w_2,\ldots,w_t)=\\
&\sum_{i=1}^n\prod_{\ell=1}^t
\biggl(\biggl(\sum_{j=1}^t (1-z_j)b_{ij}+z_j(1-b_{ij})\biggr)-w_\ell\biggr)\pmod q\,.
\end{split}
\end{equation}
Let us now take
\[
P(x)=B(I(x))\pmod q\,.
\]
To evaluate $P(x)$ at a point $x_0\in\Z_q$, a node first computes 
the vector $I(x_0)\in\Z_q^{2t}$ and then uses the evaluation formula 
\eqref{eq:b-hamming} to obtain $P(x_0)$ in time $\tilde O(nt^2)$.
We observe that $P(x)$ has degree $d\leq nt^2$. 
Furthermore, for all $i=1,2,\ldots,n$ and $h=0,1,\ldots,t$
we have
\[
\begin{split}
P\bigl(i(t+1)+h\bigr)&=
B\bigl(a_{i1},a_{i2},\ldots,a_{it},\{0,1,\ldots,t\}\setminus\{h\}
\bigr)\\
&=\bigl(\Pi_{\substack{\ell=0\\\ell\neq h}}^t(h-\ell)\bigr)c_{ih}\pmod q\,.
\end{split}
\] 
Since $0\leq c_{ih}\leq n$ and $\Pi_{\substack{\ell=0,\ell\neq h}}^t(\ell-h)$
is invertible modulo $q$, from $P\bigl(i(t+1)+h\bigr)\pmod q$ we can 
recover the integer $c_{ih}$.
From a correct proof $p_0,p_1,\ldots,p_d\in\Z_q$ 
we can by interpolation in time $\tilde O(q)$ thus recover the 
integer counts $c_{ih}$.

\subsection{A polynomial for Convolution3SUM}

This section proves Theorem~\ref{thm:ortho-hamming-conv3sum}(3).
The aim is to illustrate that we can extend simple Boolean circuits 
(a $t$-bit ripple carry adder in our case) into polynomials 
modulo $q$ and compose with input-interpolating polynomials 
to arrive at a Camelot algorithm. 

The Convolution3SUM problem (see P\v{a}tra\c{s}cu~\cite{Patrascu2010})
asks, given an array $A[1,2,\ldots,n]$ of $t$-bit integers as input,
whether there exist indices $i_1,i_2\in[n]$ with $A[i_1]+A[i_2]=A[i_1+i_2]$. 

Without loss of generality we may assume that $i_1,i_2\in[n/2]$. 
For $i\in[n/2]$, define $c_i=|\{\ell\in [n/2]:A[i]+A[\ell]=A[i+\ell]\}|$.
The sum $\sum_{i=1}^{n/2} c_i$ is the number of solutions to
the instance $A$. Thus, it suffices to define a proof polynomial
$P(x)$ from which we can recover $P(i)=c_i$ for all $i\in [n/2]$.
We will work modulo $q=\tilde O(nt^2)$. 

Let us index the $t$ bit positions of the integers in $A$ by $j=1,2,\ldots,t$, 
where $j=1$ is the least significant bit and $j=t$ is the most
significant. In time $\tilde O(nt)$ we can find $t$ univariate polynomials 
$A_1(x),A_2(x),\ldots,A_t(x)$ of degree at most $n$ such that
$A_j(i)\pmod q$ is the value of bit $j$ of $A[i]$ for all 
$i=1,2,\ldots,n$ and $j=1,2,\ldots,t$. 

In time $\tilde O(nt)$ we can evaluate a $3t$-variate
polynomial $T$
such that 
\[
T(y_1,y_2,\ldots,y_t,z_1,z_2,\ldots,z_t,w_1,w_2,\ldots,w_t)=
[y+z=w]
\]
holds for all bit-vectors%
\footnote{Caveat: Recall that $y_1$ is the {\em least} significant
bit and $y_t$ is the {\em most} significant bit by our convention.
We write $[y+z=w]$ for the $\{0,1\}$-valued indicator variable
that indicates whether $y+z=w$ holds for the $t$-bit binary
integers $y,z,w$.}
$(y_1,y_2,\ldots,y_t)\in\{0,1\}^t$,
$(z_1,z_2,\ldots,z_t)\in\{0,1\}^t$, and
$(w_1,w_2,\ldots,w_t)\in\{0,1\}^t$.
Indeed, let 
\[
\begin{split}
S(b_1,b_2,b_3)&=\\
&\!\!\!\!\!\!\!\!\!\!\!\!\!\!\!(1-b_1)(1-b_2)b_3+
(1-b_1)b_2(1-b_3)+
b_1(1-b_2)(1-b_3)+
b_1b_2b_3\\[1mm]
M(b_1,b_2,b_3)&=
(1-b_1)b_2b_3+
b_1(1-b_2)b_3+
b_1b_2(1-b_3)+
b_1b_2b_3
\end{split}
\]
be the 3-variate {\em sum} and {\em majority} polynomials, respectively, 
and define the {\em ripple carry} polynomials by the recurrence
\begin{equation}
\label{eq:ripple-carry}
c_0=0,\qquad c_j=M(y_j,z_j,c_{j-1})
\end{equation}
for $j=1,2,\ldots,t$. 

For given 
$(y_1,y_2,\ldots,y_t)\in\Z_q^t$,
$(z_1,z_2,\ldots,z_t)\in\Z_q^t$, and
$(w_1,w_2,\ldots,w_t)\in\Z_q^t$
we can now evaluate $T$ via the formula
\begin{equation}
\label{eq:t-sum}
\begin{split}
T(&y_1,y_2,\ldots,y_t,z_1,z_2,\ldots,z_t,w_1,w_2,\ldots,w_t)=\\
&(1-c_t)
\prod_{j=1}^t 
\bigl((1-w_j)(1-S(y_j,z_j,c_{j-1}))+w_jS(y_j,z_j,c_{j-1})\bigr)\,.
\end{split}
\end{equation}
This takes $O(t^2)$ operations if we use the
recurrence \eqref{eq:ripple-carry} to evaluate the 
carries $c_j$. We observe that $T$ is a 
$3t$-variate polynomial of degree $O(t^2)$.

Let us now define the proof polynomial
\[
P(x)=\sum_{\ell=1}^{n/2}T(A(x),A(\ell),A(x+\ell))\pmod q\,.
\]
We observe that $P(x)$ has degree $d=O(nt^2)$.
To evaluate $P(x)$ at a point $x_0\in\Z_q$, 
a node first computes the vector $A(x_0)\in\Z_q^t$ and
the vectors $A(x_0+\ell)\in\Z_q^t$ for each $\ell\in[n/2]$.
This takes time $\tilde O(nt)$.
The node then uses the evaluation formula \eqref{eq:t-sum} 
to obtain $P(x_0)$ in time $\tilde O(nt^2)$.
Furthermore, for all $i=1,2,\ldots,n$ we have
\[
P(i)=c_i\pmod q\,.
\]
Since $0\leq c_i\leq n/2$ and $q\geq n$, we can recover
$c_i$ from $P(i)\pmod q$.

\subsection{A polynomial for the permanent}

This polynomial illustrates the use of inclusion--exclusion formulas
in designing proof polynomials.
We prove Theorem~\ref{thm:cnf-per-hp}(2,3). 
Our starting point is Ryser's formula~\cite{Ryser1963} 
\[
\per A = 
\sum_{S\subseteq [n]}(-1)^{n-|S|}\prod_{i=1}^n\sum_{j\in S}a_{ij} 
\]
for the permanent of an $n\times n$ integer matrix $A=(a_{ij})$. 

Let us define the proof polynomial $P(x)$. We work modulo multiple
distinct primes $q$ with $q=O^*(2^{n/2})$.
Such primes can be found in time $O^*(1)$ \cite{Agrawal2004}.
Fix one such prime $q$.
In time $O^*(2^{n/2})$ we can find $n/2$ univariate polynomials 
$D_1(x),D_2(x),\ldots,D_{n/2}(x)$ of degree at most $2^{n/2}$ such 
that the vector of values
\begin{equation}
\label{eq:d-permanent}
D(x)=(D_1(x),D_2(x),\ldots,D_{n/2}(x))
\end{equation}
ranges over all the distinct vectors in $\{0,1\}^{n/2}$
modulo $q$ as $x=0,1,\ldots,2^{n/2}-1$.
Define the $n/2$-variate polynomial 
\begin{equation}
\label{eq:b-permanent}
\begin{split}
Q(z_1,&z_2,\ldots,z_{n/2})=\\
&\sum_{z_{n/2+1},z_{n/2+2},\ldots,z_{n}\in\{0,1\}}
(-1)^n
\prod_{j=1}^{n}(1-2z_j)
\prod_{i=1}^n
\sum_{j=1}^{n}
a_{ij}z_j\pmod{q}\,.
\end{split}
\end{equation}
Observe that $Q$ has degree at most $n$. Let us now take
\[
P(x)=Q(D(x))\,.
\]
To evaluate $P(x)$ at a point $x_0\in\Z_q$, 
a node first computes the vector $D(x_0)\in\Z_q^{n/2}$ and 
then uses the evaluation formula \eqref{eq:b-permanent} 
to obtain $P(x_0)$ in time $O^*(2^{n/2})$.
We observe that $P(x)$ has degree $d\leq 2^{n/2}n$.
Thus it suffices to work modulo $q=O^*(2^{n/2})$ to enable fast evaluation
and error-correction. From a correct proof $p_0,p_1,\ldots,p_d\in\Z_q$ 
we can in time $O^*(q)$ recover the values $P(0),P(1),\ldots,P(q-1)\in\Z_q$.
However, recovering the permanent is slightly more technical.
From Ryser's formula, {\em over the integers} we have
\[
\per A
=\sum_{z_{1},z_{2},\ldots,z_{n/2}\in\{0,1\}}
 Q(z_1,z_2,\ldots,z_{n/2})
=\sum_{i=0}^{2^{n/2}-1}P(i)\,.
\]
Since $|\per A|\leq 2^{n}(\max_{i,j}|a_{ij}|)^n$, one of 
our chosen primes $q$ alone may not be sufficient to reconstruct
$\per A$ from $\per A\pmod q$. Since $q=O^*(2^{n/2})$, we observe
that presenting the proof modulo $O^*(1)$ distinct primes $q$
suffices for reconstruction via the Chinese Remainder Theorem.

A similar approach works for counting the number of
Hamiltonian paths in a graph~\cite{Karp1981}. (We omit the detailed proof.)

\subsection{A polynomial for counting set covers}

This section proves Theorem~\ref{thm:set-covers}.
Let us look at another example how to work with inclusion--exclusion formulas.
Besides illustrating the use of the method, 
this example also serves to motivate why our more elaborate designs
to work with larger set families are justified.

The set cover counting problem asks, given a set $\mathcal{F}$ of subsets
of $[n]$ and an integer $0\leq t\leq n$ as input,
for the number $c_t(\mathcal{F})$ 
of $t$-tuples $(X_1,X_2,\ldots,X_t)\in\mathcal{F}^k$ 
with $X_1\cup X_2\cup\cdots X_t=[n]$. 
We have the inclusion--exclusion formula~\cite{BHK2009}
\[
c_t(\mathcal{F})
=\sum_{Y\subseteq [n]}(-1)^{n-|Y|}
|\{X\in\mathcal{F}:X\subseteq Y\}|^t\,.
\]
Let us now define the proof polynomial $P(x)$.
Again we will be working with multiple primes $q$ with $q=O^*(2^{n/2})$.
Recall the vector of polynomials $D(x)$ defined in \eqref{eq:d-permanent}. 
Define the $n/2$-variate polynomial 
\begin{equation}
\label{eq:f-t}
\begin{split}
&F_t(y_1,y_2,\ldots,y_{n/2})=\\
&\,\sum_{y_{n/2+1},y_{n/2+2},\ldots,y_{n}\in\{0,1\}}
\!\!\!\!\!\!\!\!\!\!\!(-1)^n
\prod_{j=1}^{n}(1-2y_j)
\biggl(
\sum_{X\in\mathcal{F}}
\prod_{j=1}^n
\bigl(1-(1-y_j)[j\in X]\bigr)
\biggr)^t\\
&\qquad\qquad\qquad\qquad\qquad\qquad\qquad\qquad\qquad\qquad\qquad\qquad\pmod q\,.
\end{split}
\end{equation}
We observe that the polynomial $F_t$ has degree at most $(1+t)n/2$. 
Let us now take
\begin{equation}
\label{eq:cover-proof}
P(x)=F_t(D(x))\,.
\end{equation}
To evaluate $P(x)$ at a point $x_0\in\Z_q$, 
a node first computes the vector $D(x_0)\in\Z_q^{n/2}$ and 
then uses the evaluation formula \eqref{eq:f-t} 
to obtain $P(x_0)$ in time $O^*(2^{n/2})$.
We observe that $P(x)$ has degree $d\leq 2^{n/2}(1+t)n/2$.
Thus it suffices to work modulo $q=O^*(2^{n/2})$ to enable fast evaluation
and error-correction. From a correct proof $p_0,p_1,\ldots,p_d\in\Z_q$ 
we can in time $O^*(q)$ recover the values $P(0),P(1),\ldots,P(q-1)\in\Z_q$.
From the inclusion--exclusion formula, {\em over the integers} we have
\[
c_t(\mathcal{F})
=\sum_{y_{1},y_{2},\ldots,y_{n/2}\in\{0,1\}}
 F_t(y_1,y_2,\ldots,y_{n/2})
=\sum_{i=0}^{2^{n/2}-1}P(i)\,.
\]
Since $|c_t(\mathcal{F})|\leq 2^{n(t+1)}$ and $q=O^*(2^{n/2})$, we observe
that presenting the proof modulo $O^*(1)$ distinct primes $q$
suffices for reconstruction via the Chinese Remainder Theorem.

Observe that the time to evaluate $F_t$ at any
fixed point $(y_{1},y_{2},\ldots,y_{n/2})\in\{0,1\}^{n/2}$
using \eqref{eq:f-t} is $O^*(2^{n/2}|\mathcal{F}|)$, 
so the running time at each node is $O^*(2^{n/2})$
if we assume $|\mathcal{F}|$ is bounded by a polynomial in $n$.

\medskip
\noindent
{\em Remarks.} The explicit sum over $X\in\mathcal{F}$ in \eqref{eq:f-t}
forces bad scaling when $\mathcal{F}$ is large. Assuming $\mathcal{F}$
is implicitly defined, in certain cases 
(for example, suppose that $\mathcal{F}$ is the
set of independent sets of an $n$-vertex graph $G$) we can execute the
sum implicitly, as we will do in \S\ref{sect:chromatic}
and \S\ref{sect:tutte} for the chromatic and Tutte polynomials. 
However, it appears that such implicit summation 
requires also a considerable change in the structure of the proof
polynomial $P(x)$. Indeed, contrast \eqref{eq:cover-proof} 
with \eqref{eq:part-proof}. Furthermore, executing the summation
in implicit form relies on Yates's algorithm for the chromatic
polynomial, and on a combination of Yates's algorithm and fast
matrix multiplication for the Tutte polynomial.

\section{Further applications}

\label{sect:further-appl}

This section illustrates further applications of our main results.

\subsection{Enumerating variable assignments to 2-constraints by weight}

This section proves Theorem~\ref{thm:enum-2csp} and illustrates
a further application of the $\binom{6}{2}$-linear form. 
Here it should be noted that only our use of the 
$\binom{6}{2}$-linear form is novel; the algebraic embedding 
is due to Williams~\cite{Williams2005}.

Let $z_1,z_2,\ldots,z_n$ be variables that take values over
an alphabet of $\sigma$ symbols. For convenience, let us assume
that $6$ divides $n$. Partition the variables into 
$6$ sets $Z_1,Z_2,\ldots,Z_n$ of size $n/6$ each. 

Let $\varphi_1,\varphi_2,\ldots,\varphi_m$
be constraints of arity $2$, that is, whether each constraint 
$\varphi_j$ is satisfied can be determined by looking at the 
values of exactly $2$ of the variables. For a constraint $\varphi_j$,
let us write $x_{j_1}$ and $x_{j_2}$ with $j_1<j_2$ for
these variables. 

We say that the constraint $\varphi_j$ has {\em type} $(s,t)$ 
for $1\leq s<t\leq 6$ if $(s,t)$ is the lexicographically
least pair with the property that $x_{j_1},x_{j_2}\in Z_s\cup Z_t$. 
Observe in particular that each constraint has a unique type.

For $1\leq s<t\leq 6$,
let $a_s$ and $a_t$ be assignments of values to the variables 
in $Z_s$ and $Z_t$. Let us write $f^{(s,t)}(a_s,a_t)$ 
for the number of constraints of type $(s,t)$ that are
satisfied by $a_s,a_t$. 

Let $N=\sigma^{n/6}$. That is, $N$ is the number of distinct
assignments to the variables in $Z_s$, $s=1,2,\ldots,6$.
Let $w$ be a polynomial indeterminate and define
for each $(s,t)$ with $1\leq s<t\leq 6$ the $N\times N$ 
matrix $\chi^{(s,t)}(w)$ with entries
\[
\chi^{(s,t)}_{a_s,a_t}(w)=w^{f^{(s,t)}(a_s,a_t)}\,.
\]
(Here the rows and columns of $\chi^{(s,t)}$ are indexed by
the assignments of values to the variables in $Z_s$ and $Z_t$,
respectively.) 
Observe that constructing these matrices takes $O^*(2^{2n/6})$ time and space.

Now let us study the $\binom{6}{2}$-linear form 
over the $\binom{6}{2}$ matrices $\chi^{(s,t)}(w)$. 
Viewing the form as a polynomial in $w$, we have
\[
\begin{split}
X_{\binom{6}{2}}(w)
&=\sum_{a_1,a_2,\ldots,a_6}
\prod_{1\leq s<t\leq 6}\chi^{(s,t)}_{a_s,a_t}(w)\\
&=\sum_{a_1,a_2,\ldots,a_6}
w^{\sum_{1\leq s<t\leq 6}f^{(s,t)}(a_s,a_t)}
\,.
\end{split}
\]
That is, $X_{\binom{6}{2}}(w)$ is a polynomial of degree at most $m$
with the property that the integer coefficient of each monomial $w^k$ is 
precisely the number of assignments $(a_1,a_2,\ldots,a_6)$ to the $n$ 
variables that satisfy exactly $k$ of the $m$ constraints. 
Thus, it suffices to construct $X_{\binom{6}{2}}(w)$ over the integers.
We will accomplish this by performing evaluations of $X_{\binom{6}{2}}(w)$ 
at $m+1$ distinct integer points $w_0$ and then interpolating over the
integers. For $w_0=0,1,\ldots,m$ we observe that
$0\leq X_{\binom{6}{2}}(w_0)\leq \sigma^n m^m$, so it suffices to
use the proof polynomial and the evaluation algorithm
for $X_{\binom{6}{2}}$ in \S\ref{sect:six-cliques} with $N=\sigma^{n/6}$ 
and $O^*(1)$ distinct primes $q$ to recover the integer
$X_{\binom{6}{2}}(w_0)$ via the Chinese Remainder Theorem. 

\end{document}